  \documentclass{EngC}

  \usepackage[rightcaption,raggedright]{sidecap}
  \usepackage{framed}         
  \usepackage{soul}           

  \usepackage{rotating}
  \usepackage{floatpag}
  \rotfloatpagestyle{empty}

 \usepackage{amsmath}
  \usepackage{amsthm}
  \usepackage{graphicx}

 \usepackage{makeidx}
 \makeindex

  \newcommand\cambridge{EngC}

  \theoremstyle{plain}
  \newtheorem{theorem}{Theorem}[chapter]

  \theoremstyle{definition}
  \newtheorem{definition}[theorem]{Definition}

  \theoremstyle{remark}

\usepackage{pgf,tikz}
\usepackage{pgfplots}
\usetikzlibrary{calc,shadows}
\usetikzlibrary{pgfplots.groupplots}
\usepackage{subfigure}
\usepackage{url}
\usetikzlibrary{pgfplots.groupplots}

\newtheorem{proposition}{Proposition}
  \definecolor{DarkBlue}{rgb}{0,0,0.7} 

\newcommand{\abs}[1]{ \left| #1 \right| }

\newcommand{\K}{{K}}

\newcommand{\derd}{\partial}

\usepackage{rays_defs_13}

\newcommand{\setA}{\mathcal A}

\renewcommand{\L}{L}

\newcommand{\N}{N}
\renewcommand{\S}{S}

\newcommand{\FK}{F} 

\newcommand{\sign}{\mathrm{sign}}

\renewcommand{\u}{u} 

\newcommand{\Tint}{T}
\newcommand{\Bint}{B}
\newcommand{\tauc}{\bar \tau}
\newcommand{\nuc}{\bar \nu}

\renewcommand{\d}{d}

\newcommand{\minlet}[1]{\tilde #1} 

\newcommand\SRF{\mathrm{SRF}}
\newcommand{\atom}{\vf}

\newcommand\taum{\tau_{\max}}
\newcommand\num{\nu_{\max}}

\newcommand{\mAA}{\mA}

\newcommand{\AN}{\mathrm{AN}}

\newcommand{\prsig}[0]{x}  

\newcommand{\sfunc}{s} 

\newcommand{\x}{\tau}
\newcommand{\y}{\nu}

\begin{document}
  \title[Subtitle, if you have one]
    {LaTeX2e guide for authors using the \cambridge\ design}

  \author{ALI WOOLLATT\\[3\baselineskip]
    This guide was compiled using \hbox{\cambridge.cls \version}\\[\baselineskip]
    The latest version can be downloaded from:
    https://authornet.cambridge.org/information/productionguide/
      LaTeX\_files/\cambridge.zip}


  \mainmatter


  \chapterauthor{Reinhard Heckel
    \affil{Department of Electrical and Computer Engineering, Rice University}}

\setcounter{chapter}{6}
\chapter{Super-resolution radar imaging via convex optimization}

A radar system emits probing signals and records the reflections.
Estimating the relative angles, delays, and Doppler shifts from the received signals allows to determine the locations and velocities of objects. 
However, due to practical constraints, the probing signals have finite bandwidth $B$, the received signals are observed over a finite time interval of length $T$ only, and a radar typically has only one or a few transmit and receive antennas. 
These constraints fundamentally limit the resolution up to which objects can be distinguished. 
Specifically, a radar can not distinguish objects with delay and Doppler shifts much closer than $1/B$ and $1/T$, respectively, and a radar system with $N_T$ transmit and $N_R$ receive antennas cannot distinguish objects with angels closer than $1/(N_T N_R)$. 
As a consequence, the delay, Doppler, and angular resolution of standard radars is proportional to $1/B$ and $1/T$, and $1/(N_T N_R)$. 
In this chapter, we show that the \emph{continuous} angle-delay-Doppler triplets and the corresponding attenuation factors can be resolved at much finer resolution, using ideas from compressive sensing. 
Specifically, provided the angle-delay-Doppler triplets are \emph{separated} either by factors proportional to $1/(N_T N_R-1)$ in angle, $1/B$ in delay, or $1/T$ in Doppler direction, they can be recovered a significantly smaller scale or higher resolution. 

\section{Introduction}

A traditional Single-Input Single-Output (SISO) pulse-Doppler radar system transmits a probing signal and receives the reflections from objects with a single antenna. 
By estimating the induced delays and Doppler shifts the radar system can determine the relative distances and velocities of the objects. 
However---like for any imaging system---physics imposes a limit on how well objects can be resolved. 
The resolution of a radar system is determined by the  bandwidth $B$ of the probing signals and the time interval $T$ over which the responses are observed. 
Specifically, the delay and Doppler resolution is proportional to $1/B$ and $1/T$, meaning that objects closer than that are essentially impossible to distinguish under noise. 
Since both $B$ and $T$ cannot be made arbitrarily large due to physical limitations, those two constraints fundamentally limit the resolution of a SISO radar system. 
In contrast to SISO radar systems, Mulitple-Input, Multiple-Output (MIMO) radar systems \cite{bliss_multiple-input_2003,li_mimo_2007} use multiple antennas to transmit probing signals simultaneously and record the reflections from the objects with multiple receive antennas. 
A MIMO radar can thereby, in principle, resolve the relative angles in addition to the relative distances and velocities of objects with a single measurement.
However, the angular resolution of a MIMO radar is $1/(N_T N_R)$, where $N_T$ and $N_R$ are the number of transmit and receive antennas, and is thus again limited by a physical constraint, namely the number of antennas (see Section~\ref{sec:MIMO} for a detailed argument on the resolution).

Even though objects that are simultaneously much closer than $1/(N_T N_R)$ in angle, $1/B$ in delay, and $1/T$ in Doppler direction, 
are impossible to distinguish for real world radar systems in general, it is still possible to determine the locations of the objects up to a much higher accuracy than the resolution limit of $(1/(N_T N_R),1/B,1/T)$. 
In this chapter, we discuss signal recovery techniques for building super-resolution radar systems that can achieve localization accuracy below the resolution limit. 
In more detail, we study the problem of recovering the \emph{continuous} delays and Doppler shifts in a SISO radar system, and the problem of recovering the angles, delays and Doppler shifts in a MIMO radar system, in both cases from the responses to known and suitably selected probing signals. 
As we see later, those problems---termed the super-resolution radar and super-resolution MIMO radar problems---amount to recover a signal that is sparse in a \emph{continuous} dictionary from linear measurements, and can thus be viewed as a generalization of the traditional compressive sensing problem. 

If the objects may be assumed to lie on a sufficiently \emph{coarse} grid, 
compressed sensing \cite{candes_robust_2006} based  approaches provably recover the delay-Doppler pairs for SISO radar system~\cite{herman_high-resolution_2009,baraniuk_compressive_2007,heckel_identification_2013}, and the angle-delay-Doppler triplets for MIMO radar systems~\cite{dorsch_refined_2015,strohmer_adventures_2015}. 
However, to establish those results, the aforementioned  papers assume that angles, delays, and Doppler shifts lie on a sufficiently \emph{coarse} grid, specifically a grid with spacing $1/(N_TN_R), 1/B$, and $1/T$, in angle, delay, and Doppler direction, respectively (see Section~\ref{sec:ongrid}). 
Since $N_T,N_R, B$, and $T$ are 
physical problem parameters, they can in general not be made (arbitrarily) large in order to make the grid finer. 
In fact, the coarseness of the grid is required for the measurement matrix to be incoherent, therefore the aforementioned results cannot straightforwardly be extended to a grid with significantly finer spacing. 
In some special cases, however, off-the-grid recovery is possible with standard spectral estimation techniques. 
For example, for a single input antenna and either known and constant delays (see Section~\ref{sec:redsupres}), or known and constant Doppler shifts, the super-resolution radar problem reduces to a standard 2D line spectral estimation problem \cite[Sec.~5]{strohmer_adventures_2015}. 
 For these special cases, the object locations can be recovered---off the grid---with standard spectral estimation techniques such as Prony's method, MUSIC, and ESPRIT \cite{stoica_spectral_2005}. 
In general, however, the super-resolution radar problems cannot be reduced to the classical line spectral estimation problem. Therefore, traditional spectral estimation techniques are not directly applicable. 

Recently, an alternative, convex optimization based approach to solve the classical \emph{line spectral estimation} has been proposed that is much more generally applicable than traditional line spectral estimation techniques. 
Specifically, the paper~\cite{candes_towards_2014} shows that the frequency parameters, which are the unknowns in the line spectral estimation problem, can be perfectly recovered by solving a convex total-variation norm minimization program, provided they are sufficiently separated. 
Related convex programs have been studied for compressive sensing off the grid \cite{tang_compressed_2013}, denoising \cite{bhaskar_atomic_2012}, 
signal recovery from short-time Fourier measurements \cite{aubel_theory_2015} and the SISO and MIMO super-resolution radar problems \cite{heckel_super-resolution_2015,heckel_super-resolution_2016}, and the generalized line-spectral estimation problem~\cite{heckel_generalized_2017}. 
The focus of this chapter is on explaining how this  convex optimization based approach enables high resolution in radar. In particular we discuss the results in~\cite{heckel_super-resolution_2015,heckel_super-resolution_2016}, showing that a convex program recovers the \emph{continuous} angles, delays, and Doppler shifts perfectly, provided that they are sufficiently separated. 
Furthermore, we show that a simple convex $\ell_1$-minimization program recovers the angles and delay-Doppler shifts on an \emph{arbitrarily fine} grid, again provided they are sufficiently separated. 
Finally, we provide numerical results demonstrating robustness to noise. 

\paragraph{Outline: } The remainder of this chapter is organized as follows.
In the first part we consider the SISO radar model. In more detail, Section~\ref{sec:siso} contains the radar model and formal problem statement, in Sections~\ref{sec:atomicnmin} and \ref{sec:mainares} we present the convex optimization based recovery approach and corresponding performance guarantees, and in Section~\ref{sec:discretesuperes} we show that $\ell_1$-minimization recovers the locations on an arbitrarily \emph{fine} grid. In Section~\ref{sec:numres}, we provide numerical results demonstrating that the approach is robust to noise and in Section~\ref{sec:proofoutline} we outline the proof of the main technical statements. 
In the second part, Section~\ref{sec:MIMO}, we explain how the results for SISO radar can be extended to the MIMO case. 
We conclude in Section~\ref{sec:discuss} with a discussion on challenges in applying those ideas in practice and current and future research directions.


\section{\label{sec:siso}Signal model and problem statement}

A radar system with a single transmit and single receive antenna\index{SISO} is typically modeled as a linear system. The response $y$ recorded at the receive antenna, to a probing signal, $x$, emitted at the transmit antenna, is a weighted superposition of delayed and Doppler-shifted versions of the probing signal $x$:
\begin{equation}
y(t) = \iint   \sfunc (\tau,\nu) \prsig(t-\tau)  e^{i2\pi \nu t}  d\nu d\tau.
\label{eq:ltvsys}
\end{equation}
Here, $\sfunc$ denotes the spreading function which describes the scene being sensed and $\tau$ and $\nu$ are the delays and Doppler shifts. 
Often, the moving objects are modeled by point scatterers. Mathematically, this means that the spreading function specializes to 
\[
\sfunc(\tau,\nu) = \sum_{j=1}^{\S} b_j \delta(\tau-\tauc_j) \delta(\nu-\nuc_j).
\label{eq:origradarspreadfunc}
\]
Here, $b_j$ is the (complex-valued) attenuation factor associated with the delay-Doppler pair $(\tauc_j,\nuc_j)$. 
With the spreading function above, the input-output relation \eqref{eq:ltvsys} reduces to 
\begin{align} 
y(t)
=
\sum_{j=1}^\S  b_j \prsig(t-\tauc_j)  e^{i2\pi \nuc_j t}.
\label{eq:iorelintro}
\end{align}
Thus, the received signal is a superposition of the reflections of the probing signal by the point scatterers. 
The relative distances and velocities of the $\S$-many objects can be trivially obtained from the delay-Doppler pairs $(\tauc_j,\nuc_j)$. 
In order to locate the objects, we therefore need to estimate the delay-Doppler pairs and the corresponding attenuation factors $b_j$ from a single input-output measurement, i.e., from the response $y$ to a known and suitably selected probing signal $x$. 
As we see later, the particular choice of the probing signal is crucial for good localization performance.

\subsection{\label{sec:bandtimelim} Band- and time-limitation and resolution}
\index{resolution}
The probing signal $x$ can be controlled by the system engineer and is known. 
However, due to practical and technological constraints,
it must be band-limited and approximately time-limited. Also, again due to practical constraints, we can only observe the response $y$ over a finite time interval. 
For concreteness, we assume that 
\begin{enumerate}
\item[i] we observe the response $y$ over an interval of length $T$ and that 
\item[ii]
$x$ has bandwidth $B$ and is approximately supported on a time interval of length proportional to $T$. 
\end{enumerate}
The time- and band-limitation determines the ``natural'' resolution of the system, which is the accuracy up to which the delay-Doppler pairs can be identified.
A standard pulse-Doppler radar that samples the received signal at its Nyqist rate, and performs digital matched filtering, estimates the parameters up to accuracy $1/\Bint$ and $1/\Tint$ in delay ($\tau$) and Doppler ($\nu$) directions, respectively, and therefore only identifies the delay-Doppler pairs up to the natural solution. 

From the input-output relation~\eqref{eq:iorelintro}, it is evident that band- and approximate time-limitation of the input signal $x$ implies that the response $y$ is band- and approximately time-limited as well---provided that the delay-Doppler pairs are compactly supported. In radar, due to path loss and finite velocity of the objects in the scene this is indeed the case \cite{strohmer_pseudodifferential_2006}. 
Throughout, we will therefore assume that the delay-Doppler pairs $(\tauc_j, \nuc_j)$ lie in the region 
\[
[-T/2,T/2]\times[-B/2,B/2].
\]
 This is not a restrictive assumption as this region can have area $BT \gg 1$, 
which is typically very large. 
In fact, for many applications, it is reasonable to assume that the delay-Doppler pairs lie in a region of area significantly smaller than one \cite{taubock_compressive_2010,bajwa_learning_2008,bajwa_identification_2011}, an assumption often referred to as the linear system being ``underspread''. We do not make or require this assumption here. 

By the $2WT$-Theorem \cite{slepian_bandwidth_1976},
band- and approximate time-limitation of the response $y$ implies that $y$ is essentially characterized by on the order of $BT$ coefficients. 
We therefore sample $y$ in the interval $[-T/2, T/2]$ at rate $1/B$, so as to collect $\L \defeq BT$ samples, denoted by $y_p \defeq y(p/B)$ (for simplicity we assume in the following that $\L = BT$ is an odd integer). 
As detailed in~\cite[Sec.~5]{heckel_super-resolution_2015}, those samples are given by 
\begin{align}
y_p
&= 
\sum_{j=1}^{\S} b_j  
[\mc F_{\nu_j}
\mc T_{\tau_j}
\vx ]_p
%
, \quad p = -\N,...,\N, \quad 
N \defeq \frac{L-1}{2},
\label{eq:periorel}
\end{align}
where 
\begin{align}
[\mc T_{\tau} \vx ]_p
\defeq
\frac{1}{L}
\sum_{k=-\N}^{\N} 
\left[ 
\left(
\sum_{\ell=-\N}^{\N}
x_{\ell} e^{- i2\pi \frac{\ell k}{L}  }
\right)
e^{-i2\pi k  \tau }   
\right]
e^{i2\pi \frac{p k}{L}  } 
\label{eq:deftimefreqshifts}
\end{align}
and
\[
[\mc F_{\nu} \vx ]_p
\defeq x_p e^{i2\pi p \nu }.
\]
Here, we defined the time-shifts $\tau_j \defeq  \tauc_j/T$ and frequency-shifts $\nu_j \defeq \nuc_j/B$.
To avoid ambiguity, from here onwards we refer to $(\tauc_j, \nuc_j)$ as a delay-Doppler pair and to $(\tau_j, \nu_j)$ as a time-frequency shift. 
From $(\tauc_j, \nuc_j) \in [-T/2,\allowbreak T/2] \allowbreak \times[-B/2,B/2]$ we have $(\tau_j, \nu_j) \in [-1/2,1/2]^2$. Since $\mc T_{\tau}\vx$ and $\mc F_{\nu}\vx$ are $1$-periodic in $\tau$ and $\nu$, we assume in the remainder of the chapter without loss of generality that $(\tau_j, \nu_j) \in [0,1]^2$. 
The operators $\mc T_{\tau}$ and $\mc F_{\nu}$ have an interesting interpretation as fractional time and frequency shift operators in $\complexset^\L$. In fact, if the parameters $\tau$ and $\nu$ lie on a $(1/L,1/L)$ grid, the operators $\mc F_{\nu}$ and $\mc T_{\tau}$ reduce to the ``natural'' time frequency shift operators in $\complexset^\L$, i.e., 
\[
\text{$[\mc T_{\tau} \vx ]_p = x_{p - \tau \L}$ and $[\mc F_{\nu} \vx ]_p = x_p e^{i2\pi p \nu }$.}
\]
The definition of a time shift in \eqref{eq:deftimefreqshifts} as taking the Fourier transform, 
modulating the frequency, and taking the inverse Fourier transform is a very natural definition of a \emph{continuous} time-shift $\tau_j \in [0,1]$ of a \emph{discrete} vector $\vx = \transp{[x_0,\ldots,x_{\L-1}]}$. 

Finally, note that to obtain the input-output relation~\eqref{eq:periorel} (see \cite[Sec.~5]{heckel_super-resolution_2015}) from \eqref{eq:iorelintro}, a periodic sinc function is approximated with a finite sum of sinc functions (this is where partial periodization of $x$ becomes relevant). Thus, if we take the probing signal to be essentially time-limited, then equality in~\eqref{eq:periorel} does not hold exactly. However, in
\cite[Sec.~5]{heckel_super-resolution_2015} it is shown that for a random probing signal, as considered in this chapter, the incurred relative $\ell_2$-error decays as $1/\sqrt{\L}$ 
and is therefore negligible for large $\L$. 
It is confirmed numerically in the same paper that the approximation error made in this process is negligible. 
Moreover, if we took $x$ to be $T$-periodic, then the input-output relation~\eqref{eq:periorel} becomes exact, but at the cost of the probing signal $x$ not being time-limited. 

\subsection{Formal problem statement}

From the discussion in the previous section we conclude that identification of the objects under the constraints that the probing signal $x$ is band-limited and the response $y$ to the probing signal is observed over a finite time interval, reduces to the  estimation of the triplets $\{(b_j, \tau_j, \nu_j)\}_{j=1}^S$ from the samples $\{y_p\}_{p=-N}^N$. 
Thus, in this chapter, we consider the problem of recovering those triples from the samples $\{y_p\}_{p=-N}^N$ in~\eqref{eq:periorel}. 
We call this the super-resolution radar problem, as recovering the exact time-frequency shifts $\{(\tau_j,\nu_j)\}_{j=1}^\S$ ``breaks'' the natural resolution limit  of $(1/B,1/T)$ achieved by a standard pulse-Doppler radar. 

Alternatively, one can view the super-resolution radar problem as that of recovering a signal that is $\S$-sparse in the continuous dictionary of time-frequency shifts of an $\L$-periodic sequence $x_\ell$. 
In order to see this, and  to better understand the super-resolution radar problem, it is instructive to consider two special cases.

\subsection{Time-frequency shifts on a grid \label{sec:ongrid}}

Suppose the delay-Doppler pairs $(\tauc_j, \nuc_j)$ lie on a $(\frac{1}{B},  \frac{1}{T})$-grid. As a consequence the time-frequency shifts $(\tau_j, \nu_j)$ lie on a $(\frac{1}{L},\frac{1}{L})$-grid, which in turn implies that $\tau_j \L$ and $\nu_j\L$ are integers in $\{0,\ldots,\L-1\}$. 
Thus, the super-resolution radar problem reduces to a sparse signal recovery problem with a Gabor measurement matrix. 
To see this, note that under the aforementioned assumption, the input-output relation~\eqref{eq:periorel} reduces to 
\[
y_p
= \sum_{j=1}^{\S} b_j
x_{p - \tau_j \L}
e^{i2\pi \frac{ (\nu_j \L)   p}{\L}  },  \quad p = -\N,...,\N.
\]
Writing this equation in vector-matrix form gives
\[
\vy = \mG_{\vx} \vb. 
\]
Here, the vector $\vy$ contains as entries the samples $y_p$, 
 $\mG_{\vx} \in \complexset^{\L \times \L^2}$ is the Gabor matrix with window $\vx$, defined by 
\begin{align}
[\mG_{\vx}]_{p, (k,\ell)} \defeq x_{p- \ell}  e^{i2\pi \frac{k p}{\L}}, \quad k,\ell, p = -\N,...,\N, 
\label{eq:defgabormtx}
\end{align}
and $\vb \in \complexset^{\L^2}$ is a sparse vector with the $j$-th non-zero entry given by $b_j$ and indexed by $(\tau_j\L, \nu_j\L)$. 
 
Thus, recovery of the triplets $\{(b_j,\tau_j,\nu_j)\}_{j=1}^\S$ amounts to recovering the $\S$-sparse vector $\vb$ from the measurement vector $\vy$. 
A---by now standard---recovery approach is to solve a convex $\ell_1$-norm-minimization program. 
From \cite[Thm.~5.1]{krahmer_suprema_2014} we know that, provided the $x_\ell$ are i.i.d.~sub-Gaussian random variables, and provided that $S\le c L/(\log L)^4$ for a sufficiently small numerical constant $c$,  with high probability, all $\S$-sparse vectors $\vb$ can be recovered from $\vy$ via $\ell_1$-minimization.
Note that the result \cite[Thm.~5.1]{krahmer_suprema_2014} only applies to the Gabor matrix $\mG_{\vx}$ and therefore does not apply to the super-resolution problem where the ``columns'' $\mc F_{\nu} \mc T_{\tau} \vx$ are highly correlated. 
In fact, the two problems are conceptually very different: \cite[Thm.~5.1]{krahmer_suprema_2014} shows that the columns of the Gabor matrix $\mG_{\vx}$ are nearly orthogonal, while the ``columns'' $\mc F_{\nu} \mc T_{\tau} \vx$ are extremely correlated for two pairs of time-frequency shifts that are close.

\subsection{Only time or only frequency shifts \label{sec:redsupres}}

Next, suppose we only have time- or only frequency shifts.
In both cases, recovery of the unknowns $\{(b_j,\tau_j)\}$ and $\{(b_j,\nu_j)\}$, respectively, is equivalent to the recovery of a weighted superposition of spikes from low-frequency samples. Specifically, if we only have frequency shifts, and therefore $\tau_j = 0$ for all $j$, the input-output relation~\eqref{eq:periorel} reduces to 
\begin{align}
y_p = 
 x_p \sum_{j=1}^{\S} b_j
e^{i2\pi p \nu_j  }, \quad p = -\N,...,\N.
\label{eq:supres}
\end{align}
Note that the samples $\{y_p\}$ above are samples of a mixture of $\S$ complex sinusoids, and  estimation of the coefficients $\{(b_j,\nu_j)\}$ corresponds to determining the magnitudes and the frequency components of these sinusoids. 
Estimating the coefficients $\{(b_j,\nu_j)\}$ from the samples $\{y_p\}$ is known as a line spectral estimation problem and can be solved using classical approaches such as Prony's method \cite[Ch.~2]{gershman_space-time_2005}, 
as well as convex programming based approaches~\cite{candes_towards_2014}. 
An analogous situation arises when there are only time shifts ($\nu_j = 0$ for all $j$) as taking the discrete Fourier transform of $y_p$ yields a relation exactly equal to equation~\eqref{eq:supres}.

\section{\label{sec:atomicnmin}Atomic norm minimization and associated performance guarantees}

We next present a convex optimization\index{convex optimization} based recovery algorithm. Even though the corresponding convex program can be solved in polynomial time, standard solvers are currently computationally very expensive, limiting the practical applicability.
However, in Section~\ref{sec:discretesuperes} we discuss a very closely related convex program that has a significantly better computational efficiency at the cost of making a small griding error that is due to a discretization step. 
Since the results and intuition for both approaches are nearly the same, we start with discussing the continuous case in this section.


\subsection{Atomic norm minimization}

\index{atomic norm}
We first define for convenience the vector $\vr_j \defeq [\tau_j,\nu_j]$, and write the input-output relation \eqref{eq:periorel} in matrix-vector form:
\begin{align}
\vy = \mG_{\vx}\herm{\mF} \vz, 
\quad \vz = \sum_{j=1}^{\S} b_j \vf(\vr_j).
\label{eq:syseqinw}
\end{align}
Here, $\herm{\mF} \in \complexset^{\L^2 \times \L^2}$ is the (inverse) 2D discrete Fourier transform matrix with the entry in the $(k,\ell)$-th row and $(r,q)$-th column given by $[\herm{\mF}]_{(k,\ell), (r,q)} \defeq \frac{1}{\L^2} e^{i2\pi \frac{qk + r\ell}{\L}}$ and the entries of the  vector $\vf$ are given by $[\vf(\vr)]_{(r,q)} \defeq e^{-i2\pi (r\tau + q \nu)}$, $k, \ell, q, r = -\N, \ldots, \N$
(here, and in the following we use for convenience a two or three dimensional index to refer to entries of vectors and matrices). 
Moreover, $\mG_{\vx} \in \complexset^{\L \times \L^2}$ is the Gabor matrix defined in equation~\eqref{eq:defgabormtx}. 

The significance of the representation in \eqref{eq:syseqinw} is that recovery of the unknowns $\{(b_j,\vr_j)\}$ from $\vz$ is a 2D line spectral estimation problem that can be solved with standard spectral estimation techniques such as Prony's method \cite{stoica_spectral_2005}. 
Therefore, we only need to recover $\vz \in \complexset^{\L^2}$ from $\vy \in \complexset^{\L}$. 
To do so, we use that $\vz$ is a sparse linear combination of  atoms in the set $\setA \defeq \{ \atom(\vr), \vr \in [0,1]^3\}$. 
A regularizer that promotes such a sparse linear combination is the atomic norm induced by these signals~\cite{chandrasekaran_convex_2012}, defined as 
\[
\norm[\setA]{\vz} 
\defeq \inf_{b_k \in \complexset, \vr_k \in [0,1]^2} \left\{ \sum_k |b_k| \colon \vz = \sum_k b_k \atom(\vr_k) \right\}.
\]
We estimate $\vz$ by solving the basis pursuit type atomic norm minimization problem problem  
\begin{align}
\AN(\vy) \colon \;\; \underset{\minlet{\vz}  }{\text{minimize}} \,  \norm[\setA]{\minlet{\vz} } \; \text{ subject to } \; \vy = \mAA \minlet{\vz}.
\label{eq:primal}
\end{align}
To summarize, we estimate the attenuation factors $b_k$ and time-frequency shifts $\vr_k$ from $\vy$ by 
\begin{enumerate}
\item[i] solving $\AN(\vy)$ in order to obtain $\vz$, 
\item[ii] estimating the $\vr_k$ from $\vz$ by solving the corresponding 2D-line spectral estimation problem, and 
\item[iii] solving the linear system of equations 
$
\vy = \sum_{k=0}^{\S-1} b_k \mAA \atom(\vr_k)
$ for the $b_k$. 
\end{enumerate}
We remark that the $\vr_k$ may be obtained more directly from a solution to the dual of \eqref{eq:primal} \cite[Sec.~6]{heckel_super-resolution_2015}, 
see also \cite[Sec.~3.1]{bhaskar_atomic_2012}, \cite[Sec.~4]{candes_towards_2014}, \cite[Sec.~2.2]{tang_compressed_2013} for details on this approach applied to related problems. 

Since computation of the atomic norm involves taking the infimum over infinitely many parameters, finding a solution to $\AN(\vy)$ may appear to be daunting. 
For the one-dimensional case (i.e., only time or frequency shifts), the atomic norm can be characterized in terms of linear matrix inequalities \cite[Prop.~2.1]{tang_compressed_2013}. This characterization is based on the Vandermonde decomposition lemma for Toeplitz matrices,  
and allows to formulate the atomic norm minimization program as a semidefinite program  that can be solved in polynomial time. 
While this lemma generalizes to higher dimensions \cite[Thm.~1]{yang_vandermonde_2015}, it fundamentally comes with a rank constraint 
that appears to prohibit an straightforward characterization of the atomic norm in terms of linear matrix inequalities. 
Nevertheless, based on \cite[Thm.~1]{yang_vandermonde_2015}, one can obtain a semidefinite programming \emph{relaxation} of $\AN(\vy)$, which can be solved in polynomial time. 
Similarly, a solution of the dual of $\AN(\vy)$ can be found with a semidefinite programming relaxation. 
Since the computational complexity of the corresponding semidefinite programs is quite large, we will not dive into the details of those semidefinite programming relaxations. 
As mentioned before, instead, we show in Section \ref{sec:discretesuperes} that the parameters $\{\vr_j\}$ can be recovered on an arbitrarily fine grid via $\ell_1$-minimization. 
While this leads to a gridding error, the grid may be chosen sufficiently fine for the gridding error to be negligible compared to the error induced by additive noise, and in practice, there is always some additive noise. 


\subsection{\label{sec:mainares}Recovery guarantees for atomic norm minimization}

We consider a random probing signal by taking the samples of the probing signal $x_\ell$ in \eqref{eq:periorel} to be i.i.d. Gaussian random variables. 
More generally, the result presented below continues to hold if we choose the samples as sub-Gaussian random variables, for example as random signs. 
Note that the probing signal can be chosen by the radar engineer, therefore, choosing the coefficients at random is not problematic and can be done in practice. 
The theorem stated below shows that, with high probability, the triplets $\{(b_j,\tau_j,\nu_j)\}$ can be recovered perfectly from the samples by solving a convex program, provided that the number of time-frequency shifts is sufficiently smaller than the number of measurement, and provided that the following minimum separation condition holds:

\begin{definition}[Minimum separation condition] We say the time-frequency shifts $(\tau_j,\nu_j) \in [0,1]^2, j = 1,\ldots,\S$, satisfy the minimum separation condition if
\begin{align}
\max(|\tau_j - \tau_{j'}|, |\nu_j - \nu_{j'}| ) \geq \frac{2.38}{\N}
, \text{ for all } j\neq j',
\label{eq:minsepcond}
\end{align}
where $|\tau_j - \tau_{j'}|$ is the wrap-around distance on the unit circle. For example, $|3/4-1/2|=1/4$ but $|5/6-1/6|=1/3\neq 2/3$. 
\end{definition}

Note that the time-frequency shifts must not be separated in both time \emph{and} frequency, for example the minimum separation condition can hold even when $\tau_j = \tau_{j'}$ for some $j\neq j'$. 
The main result on recovery via atomic norm minimization from the paper~\cite{heckel_super-resolution_2015} is stated next.

\begin{theorem} Assume that the samples of the probing signal $\vx \in \complexset^\L$ are i.i.d.~$\mathcal N(0,1/\L)$ random variables.  
Consider a signal where the sign of the attenuation factors $\{b_j\}_{j=1}^{\S}$ is i.i.d.~uniform on $\{-1,1\}$ or the complex unit disc, 
and suppose that the time-frequency shifts $\{(\tau_j,\nu_j)\}_{j=1}^{\S}$
obey the minimum separation condition. 
Furthermore, choose $\delta > 0$ and assume that the number of non-zero attenuation factors, $\S$, and the number of measurements, $\L$, obey 
\begin{align*}
S\le c\frac{L}{\log^3(L/\delta)}, 
\label{eq:llinscon}
\end{align*}
where $c$ is a numerical constant. Then, with probability at least $1-\delta$, $\vz = \sum_{j=1}^{\S} b_j \vf(\vr_j)$ is the unique minimizer of $\AN(\vy)$, $\vy = \mG_{\vx} \herm{\mF} \vz$. 
\label{thm:mainres}
\end{theorem}

This result is essentially optimal in terms of the allowed sparsity level, as the number $\S$ of unknowns can be linear---up to a logarithmic-factor---in the number of observations $L$. Even when we are given the time-frequency shifts $(\tau_j,\nu_j)$, we can only hope to recover the corresponding attenuation factors $b_j$ by solving the linear system of equations in~\eqref{eq:periorel}, provided that $\S \leq \L$. 

Since the complex-valued coefficients $b_j$ in the radar model describe the attenuation factors, it is natural to assume that the phases of different $b_j$ are independent from each other and are uniformly distributed on the unit circle of the complex plane. 
Indeed, standard models for wireless communication channels and radar~\cite{bello_characterization_1963}, assume the coefficients $\{b_j\}$ to be complex Gaussian distributed. Nevertheless, we believe that the random sign assumption is not necessary for Theorem~\ref{thm:mainres} to hold. In Section~\ref{sec:discuss}, we discuss a closely related problem which does not require the random sign assumption, and thus provides a basis for the claim of the random sign assumption not being necessary. 
Finally, we would like to point out that Theorem \ref{thm:mainres} continues to hold for sub-Gaussian sequences $x_\ell$, for example random signs.

\subsection{Necessity of minimum separation\label{sec:necessity}}

Theorem~\ref{thm:mainres} imposes a minimum-separation condition, and indeed some form of separation between the time-frequency shifts is necessary for \emph{stable} recovery. 
To be specific, we consider the simpler problem of line spectral estimation (see Section \ref{sec:redsupres}) that is obtained from our setup by setting $\tau_j=0$ for all $j$. 
Clearly, any condition necessary for the line spectral estimation problem is also necessary for the super-resolution radar problem. 

If there are $\S'$ frequencies $\{\nu_k\}_{k=1}^{S'}$ in an interval of length smaller than $\frac{2S'}{L}$, and $\S'$ is large, then in the presence of even a tiny amount of noise, stable recovery of the attenuation factors and time-frequency shifts even from 
\[
\vz = \sum_{j=1}^{\S} b_j \vf(\vr_j),
\]
where we set, with a slight abuse of notation, $[\vf(\nu_j)] = e^{i2\pi j \nu_j}$, is not possible (see \cite[Thm.~1.1]{donoho1992superresolution} and \cite[Sec.~1.7]{candes_towards_2014}). Condition~\eqref{eq:minsepcond} 
allows us to have  $0.4\,\S'$ time-frequency shifts in an interval of length $\frac{2\S'}{\L}$, which is optimal up to the constant $0.4$. 

This argument illustrates that for stable recovery, it is relevant whether a number of frequencies, say $\S$ many, cluster together in a small interval of size smaller than $\S/L$. 
To illustrate this point further, consider again the simpler problem of line spectral estimation, i.e., recovery or the frequencies from $\vz = \sum_{j=1}^{\S} b_j \vf(\nu_j)$, with $[\vf(\nu_j)] = e^{i2\pi j \nu_j}$. 
Consider the following (Vandermonde) matrix parameterized by $\S$ and $\epsilon$:
\[
\mV = [\vf(0), \vf( (1-\epsilon)/\L ), \ldots, \vf_\L( (2\S-1)(1-\epsilon)/\L ) ].
\]
We next state a theorem, which lower bounds the condition number of $\mV$. The lower bound implies that there are signals with $\S$ many frequencies in an interval smaller than $\S/L$, that are indistinguishable even under a tiny amount of additive noise. 

\begin{theorem}[{\cite[Thm.~1.3]{moitra_super-resolution}}
]
Fix some $\epsilon \in (0,1)$, let $\K = \frac{1}{1-\epsilon} \L$, and let $\S = O(\log (\L/ (1-\epsilon)))$. 
Then the matrix $\mV$ has condition number at least $e^{O(\epsilon \S)}$. 
\end{theorem}

\begin{figure}[h]
\begin{center}
\begin{tikzpicture}
\begin{groupplot}[group style={group size=3 by 1, 
horizontal sep=1cm }, width=0.4\textwidth, xticklabel
    style={/pgf/number format/fixed, /pgf/number format/precision=3},
    legend pos=outer north east,
  ]
\nextgroupplot[
no markers, 
enlarge x limits=false, 
xlabel= {$1-\epsilon$},
ylabel={1/$\kappa$},
]
 
\addplot +[no markers,black,solid] table[x index=0,y index=1]{./dat/condnumberFourier.dat};
\addlegendentry{$\S=2$}

\addplot +[no markers,black,dashed] table[x index=0,y index=2]{./dat/condnumberFourier.dat};
\addlegendentry{$\S=4$}

\addplot +[no markers,black,dotted] table[x index=0,y index=3]{./dat/condnumberFourier.dat};
\addlegendentry{$\S=8$}

\addplot +[no markers,black,dashdotted] table[x index=0,y index=4]{./dat/condnumberFourier.dat};
\addlegendentry{$\S=16$}

\addplot +[no markers,black,densely dotted] table[x index=0,y index=5]{./dat/condnumberFourier.dat};
\addlegendentry{$\S=32$}

\end{groupplot}
\end{tikzpicture}
\end{center}

\caption{\label{fig:conditionnumber}  
Inverse of the condition number $\kappa$ of the matrix $\mV$ with entries $[\mV]_{pq} = e^{-i2\pi p q (1-\epsilon) / \L }$, $\L=200$, and $q=1,\ldots, \S$, for different values of the number of sources $\S$ as a function of the separation between frequencies of $(1-\epsilon)/L$. 
}
\end{figure}
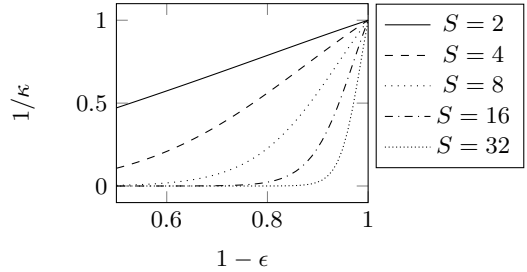

The theorem implies that there exists a vector $\vb$ with unit norm that obeys $\norm[2]{\mV \vb} \leq e^{-O(\epsilon \S)}$. As a consequence,
\[
\sum_{\ell=-\N}^{\N}
\left(
\sum_{ j \text{ odd} } b_j e^{i2\pi \nu_j \ell}
+
\sum_{j \text{ even}} b_j e^{i2\pi \nu_j \ell}
\right)^2
= 
\norm[2]{\mV \vb}^2 
\leq 
e^{-O(\epsilon \S)},
\]
which means that there are two sets of $\S$ many point sources each, with separation $\frac{2(1-\epsilon)}{\L}$, but telling them apart requires an exponentially small additive error. 
To obtain intuition on the constants involved in the statement, we plot the condition number of $\mV$ for different values of $\S$
and $\epsilon$ in Figure~\ref{fig:conditionnumber}. 

While those two arguments show that some form of separation between the time-frequency shifts is necessary, the exact form of separation required in \eqref{eq:minsepcond} may not be necessary for stable recovery and less restrictive conditions may suffice.
Indeed, in the simpler problem of line spectral estimation, Donoho~\cite{donoho1992superresolution} showed that stable super-resolution is possible via an exhaustive search algorithm even when condition~\eqref{eq:minsepcond} is violated locally, as long as every interval of the $\nu$-axis of length $\frac{2\S'}{\L}$ contains less than $\S'/2$ frequencies and $\S'$ is small (in practice, think of $\S'\lesssim 10$). 




\subsection{Implications for the detection accuracy of radar systems}

Translated to the continuous setup, Theorem~\ref{thm:mainres} implies that with high probability, the triplets $(b_j, \tauc_j,\nuc_j)$ can be identified perfectly provided that  
\begin{equation}
	\label{eq:mindist}
|\tauc_j -\tauc_{j'}| \geq \frac{4.77}{B} \,\text { or } \, |\nuc_j - \nuc_{j'}| \geq \frac{4.77}{T}, \quad \text{ for all } j\neq j',
\end{equation}
and provided that $S\le c\frac{BT}{\log^3 (BT)}$. 
Since we can \emph{exactly} identify the delay-Doppler pairs $(\tauc_j,\nuc_j)$, as opposed to only localizing them on a grid, this result offers a significant improvement in resolution over conventional radar techniques. 
Specifically, with a standard pulse-Doppler radar which samples the received signal and performs digital matched filtering in order to detect the objects, 
the delay-Dopper shifts $(\tauc_j,\nuc_j)$ can only be estimated up to an uncertainty of about $(1/T,1/B)$. 

We hasten to add that in the radar literature, the term super-resolution is often used for  the ability to resolve objects that are very close---even closer than the Rayleigh resolution limit \cite{quinquis_radar_2004} that is proportional to $1/B$ and $1/T$ for delay and Doppler resolution, respectively.  
The norm minimization approach discussed here permits identification of \emph{each} object with precision much higher than $1/B$ and $1/T$ as long as the other objects are not too close, specifically other objects should be separated by a constant multiple of the Rayleigh resolution limit as formalized by the minimum separation condition~\eqref{eq:mindist}. 
Recall that, however, any method that attempts to recover objects closer than the resolution limit can only succeed if there are very few objects below that limit, since resolving many objects that are all below the resolution limit is in general impossible, as discussed previously in Section~\ref{sec:necessity}.

Finally, recall that the approach discussed here allows the delay-Doppler pairs $(\tauc_j,\nuc_j)$ to lie in $[-T/2, T/2] \times [-B/2,B/2]$ so the delay-Doppler pairs can lie in a rectangle of area $\L=BT \gg 1$. 
The ability to handle a potentially large region in which delay-Doppler pairs can lie in is important in radar applications, since we might need to resolve objects with large relative distances and relative velocities. 

\subsection{Can standard non-parametric methods yield similar performance?}

We finally note that standard non-parametric estimation methods such as the MUSIC algorithm can in general \emph{not} be applied directly to the super-resolution radar problem. 
The reason is that MUSIC relies on \emph{multiple} measurements (often referred to as snapshots) \cite[Sec.~6.3]{stoica_spectral_2005}, whereas we assume only a \emph{single} measurement $\{y_p\}_{p = -\N}^\N$ to be available.
In our context, multiple measurements would amount to carrying out multiple, independent input-output measurements.
However, by choosing the probing signal $\vx$ in \eqref{eq:periorel} to be periodic, a single measurement  can be transformed into multiple measurements and for that case, for example the MUSIC algorithm may be applied. However, this approach, discussed in detail in \cite[Appendix~H]{heckel_super-resolution_2015}, requires the frequencies $\{\nu_j\}$ to be distinct, the time-shifts $\{\tau_j\}$ to lie in a significantly smaller range than $[0,1]$, and $S < \sqrt{L}$, as opposed to the much milder condition $S < L/\log^3(L/\delta)$ required by the convex program. 
In addition, applying MUSIC in that way is (significantly) more sensitive to noise than the convex programming based approach discussed in this chapter. 
If multiple measurements are available, for example by observing distinct paths of a signal by an array of antennas, the situation might be different. For that case, subspace methods have been studied for delay-Doppler estimation in the paper \cite{jakobsson_subspace-based_1998}.



\newcommand{\dT}{\mathcal S} 

\section{\label{sec:discretesuperes} Super-resolution radar on a fine grid}

An practical approach to estimate the triplets $\{(b_j,\tau_j, \nu_j)\}$ from the received signal $\vy$ in the input-output relation~\eqref{eq:periorel} is to suppose the time-frequency shifts lie on a \emph{fine} grid, and solve the problem on that grid. 
In general this leads to a gridding error, which, however, becomes small as the grid gets finer~\cite{tang_sparse_2013}.
We next discuss the corresponding (discrete) sparse signal recovery problem. 

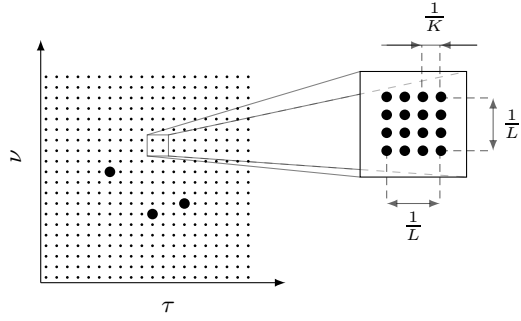
\begin{figure}
\begin{center}
	\begin{tikzpicture}[scale=1.4,>=latex] 
	\draw[->] (0,3) -- (2.3,3); 
	\draw[->] (0,3) -- (0,5.3);
	\node at (1.2,2.9) [anchor=north] {$\tau$};
	\node at (-0.1,4.2) [anchor=south, rotate=90] {$\nu$};	
	
	\foreach \x in {0.05,0.15,...,2} 
	\foreach \y in {3.05,3.15,...,5} 
	\fill [black,opacity=1] (\x,\y) circle (0.013);

	\fill [color=black,] (1.05,3.65) circle (0.05);
	\fill [color=black,] (0.65,4.05) circle (0.05);
	\fill [color=black,] (1.35,3.75) circle (0.05);

	\coordinate (a) at (3,4);
	\coordinate (b) at (4,4);
	\coordinate (c) at (4,5);
	\coordinate (d) at (3,5);
		
	\draw [opacity=0.5] (1,4.2) rectangle (1.2,4.4);
	\draw [opacity=0.5] (1,4.2) -- (a);
	\draw [opacity=0.5] (1.2,4.2) -- (b);
	\draw [opacity=0.5] (1.2,4.4) -- (c);
	\draw [opacity=0.5] (1,4.4) -- (d);
	
	\draw[fill=white] (a) rectangle (c);
	\draw [dashed,opacity=0.3] (1.2,4.2) -- (b);
	\draw [dashed,opacity=0.3] (1.2,4.4) -- (c);
	
	\foreach \x in {3.25,3.42,...,3.85} 
	\foreach \y in {4.25,4.42,...,4.85} 
	\fill [black,opacity=1] (\x,\y) circle (0.05);
	
	\draw [densely dashed,opacity=0.6] (3.25,4.25) -- (3.25,3.75);
	\draw [densely dashed,opacity=0.6] (3.75,4.25) -- (3.75,3.75);
	\draw [<->,opacity=0.6,>=latex]  (3.25,3.75) --(3.75,3.75);
	
	\node at (3.5,3.75) [anchor=north] {$\frac{1}{L}$};
	
	\draw [densely dashed,opacity=0.6] (3.75,4.25) -- (4.25,4.25);
	\draw [densely dashed,opacity=0.6] (3.75,4.75) -- (4.25,4.75);
	\draw [<->,opacity=0.6,>=latex]  (4.25,4.25) --(4.25,4.75);
	
	\node at (4.25,4.5) [anchor=west] {$\frac{1}{L}$};
	
	\draw [densely dashed,opacity=0.6] (3.75,4.85) -- (3.75,5.25);
	\draw [densely dashed,opacity=0.6] (3.58,4.85) -- (3.58,5.25);
	
	\draw [-,opacity=0.6,>=latex]  (3.25,5.25) --(4.1,5.25);
	\draw [->,opacity=0.6,>=latex]  (3.2,5.25) --(3.58,5.25);
	\draw [<-,opacity=0.6,>=latex]  (3.75,5.25) --(4.1,5.25);
	\node at (3.7,5.3) [anchor=south] {$\frac{1}{K}$};
	
\end{tikzpicture}
\end{center}
\caption{\label{fig:ongrid}
Time frequency shifts that lie on a grid: $(1/\L,1/\L)$ is the ``natural'' grid, and $(1/\K,1/\K)$ is the fine grid. Each dot corresponds to a potential non-zero, and the larger dots correspond to the actual non-zeros $\{b_j\}$. 
}
\end{figure}

Suppose the time-frequency shifts lie on a \emph{fine} grid with spacing $(1/\K,1/\K)$, where $\K$ is an integer obeying $\K \geq \L$. 
Let $\vb \in \complexset^{\K^2}$ be the signal with each entry $b_{m,n}$ corresponding to one of the grid points, with non-zeros equal to the attenuation factors $b_j$ for the time-frequency shifts $(\tau_j,\nu_j), j = 1,\ldots,\S$. 
See Figure~\ref{fig:ongrid} for an illustration. 
With this assumption, the input-output relation~\eqref{eq:periorel} becomes:
\[
y_p 
= 
\sum_{m,n=0}^{\K-1} b_{m,n}  
[\mc F_{m/K}
\mc T_{n/K}
\vx]_p, \quad p=-\N,\ldots,\N. 
\]
Writing this relation in matrix-vector form yields
\[
\vy = \mR \vb,
\]
where, as before, $\vy$ is the vector containing as entries the values $y_p$, and $\mR \in \complexset^{\L \times \K^2}$, is the matrix with $(m,n)$-th column given by $\mc F_{m/K}
\mc T_{n/K} \vx$. 
The matrix $\mR$ contains as columns ``fractional'' time-frequency shifts of the sequence $x_\ell$. If $\K = \L$, $\mR$ contains as columns only ``whole'' time-frequency shifts of $\vx$ and $\mR$ is equal to the Gabor matrix  $\mG_{\vx}$ defined by \eqref{eq:defgabormtx}. 
In this sense, $\K = \L$ is the natural grid (see~Section \ref{sec:ongrid}) and the ratio $\SRF \defeq\K/\L$ can be interpreted as a super-resolution factor. The super-resolution factor determines by how much the $(1/\K,1/\K)$ grid is finer than the original $(1/\L,1/\L)$ grid.

A standard approach to the recovery of the sparse signal $\vb$ from the underdetermined linear system of equations $\vy = \mR \vb$ is to solve the following convex program:
\begin{align}
	\mathrm{L1}(\vy) \colon \;\;
\underset{\minlet{\vb}}{\text{minimize}} \; \norm[1]{\minlet{\vb}} \text{ subject to } \vy = \mR \minlet{\vb}.
\label{eq:l1minmG}
\end{align}
The following theorem is the main result from~\cite{heckel_super-resolution_2015} for recovery on the fine grid.   
\begin{theorem} Assume that the samples of the probing signal $\vx$ are i.i.d. $\mathcal N(0,1/\L)$ random variables, $L=2N+1$, and that $\L = 2\N+1\geq 1024$. 
Consider a signal $\vb$ supported on $\mathcal{S} \subseteq \{0,...,\K-1\}^2$, and suppose that it satisfies the minimum separation condition
\[
\min_{(m,n), (m', n') \in \mathcal{S} \colon (m,n) \neq ( m', n')} 
\frac{1}{\K} \max(|m- m'|, |n - n'|) \geq \frac{2.38}{\N}.
\]
Moreover, suppose that the non-zeros of $\vb$ are i.i.d.~uniform on $\{-1,1\}$ or the complex unit disk. 
Choose $\delta>0$, let $\vy = \mR \vb$ be the measurement corresponding to $\vb$, and suppose that the number of non-zeros of $\vb$ is sufficiently smaller than the number of samples $\L$
\begin{align*}
\S\le c\frac{\L}{\log^3(\L/\delta)}, 
\end{align*}
where $c$ is a numerical constant. Then, with probability at least $1-\delta$, $\vb$ is the unique minimizer of $\mathrm{L1}(\vy)$, $\vy = \mR \vb$. 
\label{cor:discretesuperres}
\end{theorem}

Note that Theorem~\ref{cor:discretesuperres} does not impose any restriction on $\K$, in particular it can be arbitrarily large. 
The proof of Theorem \ref{cor:discretesuperres}, discussed in Section~\ref{sec:proofoutline} is closely linked to that of Theorem \ref{thm:mainres}. 

\subsection{\label{sec:impdet}Implementation details}

The matrix $\mR$ has dimension $\L \times \K^2$, thus as the grid becomes finer, i.e., $\K$ becomes larger, the complexity of solving $\mathrm{L1}(\vy)$ increases. 
The complexity of solving $\mathrm{L1}(\vy)$ can be managed as follows. First, the complexity of first-order convex optimization algorithms (such as TFOCS \cite{becker_templates_2011}) for solving $\mathrm{L1}(\vy)$ is dominated by multiplications with the matrices $\mR$ and $\herm\mR$. Due to the structure of $\mR$, those multiplications can be done very efficiently utilizing the fast Fourier transform. Second, in practice we have 
$
(\tauc_j, \nuc_j) \in [0,\taum]\times [0, \num]
$, 
which means that 
\begin{align}
(\tau_j, \nu_j) \in 
\left[0, \frac{\taum  }{ T }  \right] 
\times 
\left[0, \frac{\num }{ B}  \right] .
\label{eq:restaunun}
\end{align}
It is therefore sufficient to consider the restriction of $\mR$ to the 
$\frac{\taum \num K^2}{BT} = \taum \num \L \cdot\SRF^2$ many columns corresponding to the time-frequency shifts $(\tau_j, \nu_j)$ satisfying  \eqref{eq:restaunun}. Since typically $\taum \num \ll BT=\L$, this results in a significant reduction of the problem size.

\subsection{Numerical results and robustness\label{sec:numres}}

We next discuss numerical results which show that the convex optimization based super-resolution approach is robust to noise. 
Consider the following modification of $\ell_1$-norm minimization, which accounts for noise and the gridding error:
\[
\text{L1-ERR}\colon \underset{\minlet{\vb}}{\text{minimize}} \; \norm[1]{\minlet{\vb}} \text{ subject to } 
\norm[2]{\vy - \mR \minlet{\vb} }^2 \leq \delta.
\]
The parameter $\delta$ is chosen on the order of the noise variance or the expected gridding error. 

The paper~\cite{heckel_super-resolution_2015} considered a synthetic problem with $\L = 201$, where each problem instance is generated by drawing $S=10$ 
time-frequency shifts $(\tau_j,\nu_j)$ uniformly at random from the interval $[0,2/\sqrt{201}]^2$. This amounts to drawing the corresponding delay-Doppler pairs $(\tauc_j,\nuc_j)$ from the interval $[0,2]\times [0,2]$. 
The attenuation factors $b_j$ corresponding to the time-frequency shifts were drawn uniformly at random from the complex unit disc, independently across $j$. 
Measurements were then obtained according to the input-output relation~\eqref{eq:periorel}. 

Figure \ref{fig:realrecov} depicts the average resolution error versus the super-resolution factor $\SRF = K/L$. The resolution error is defined as the average over $j=1,\ldots,S$ of $\L \sqrt{ (\hat \tau_j - \tau_j)^2 + (\hat \nu_j - \nu_j)^2}$, where the $(\hat \tau_j, \hat \nu_j)$ are the estimates of the time-frequency shifts extracted from a solution of \text{L1-ERR}, obtained with the SPGL1 solver \cite{BergFriedlander:2008}. 
Note that the resolution attained at $\SRF=1$ corresponds to the resolution attained by matched filtering and by the compressive sensing radar architecture~\cite{herman_high-resolution_2009} discussed in Section \ref{sec:ongrid}. 

As mentioned before, there are two error sources incurred by this approach. 
The first is the gridding error obtained by assuming the points lie on a fine grid with grid constant $(1/\K,1/\K)$, which decays in $\K$. 
The second is the additive noise error, which is constant. 
The figure shows that for $\SRF$ larger than one, the resolution is significantly improved using the super-resolution radar approach. 
Moreover we see that for small super-resolution factors $\SRF$, the gridding error dominates, while for large values of $\SRF$, the additive noise error dominates. 
In this experiment, the gridding error approximately decays as $1/\SRF$. 
The experiment demonstrates that in practice solving the super-resolution radar problem on a fine grid is essentially as good as solving it on the continuum---provided the super-resolution factor is chosen sufficiently large, so that the gridding error becomes negligible relative to the error due to additive noise.

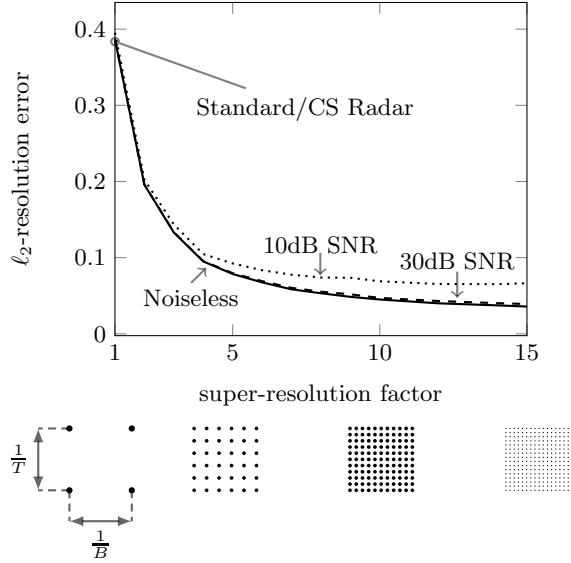
\begin{figure}
\begin{center}
 \begin{tikzpicture}[scale=1,thick] 
 \begin{axis}[xlabel={super-resolution factor}, ylabel = {$\ell_2$-resolution error}, 
 xtick={1,5,10,15,20}, 
  xmin =1,xmax=15, ymax=0.435, height=6cm,width=7cm]   
 \addplot +[thick,mark=none,black] table[x index=0,y index=2]{./dat/offgrid_data.dat};
\addplot +[thick,mark=none,dotted,black] table[x index=0,y index=3]{./dat/offgrid_data.dat}; 
  \addplot +[thick,mark=none,dashed,black] table[x index=0,y index=5]{./dat/offgrid_data.dat};
  \end{axis}
  
   \node (a) at (0,3.9) [circle,draw,inner sep=1pt,opacity=0.5] {};
   \node (b) at (2.5,3) {Standard/CS Radar}; 
   \draw[opacity=0.5] (a) -- (b);


\node at (1,0.5) {\textcolor{black}{Noiseless}};
 \draw[opacity=0.5,->] (1,0.7) -- (1.2 ,0.9);
{\node at (4.5,1) {\textcolor{black}{30dB SNR}};}
  \draw[opacity=0.5,->] (4.5,0.85) -- (4.5,0.5);
{\node at (2.7,1.2) {\textcolor{black}{10dB SNR}};} 
 \draw[opacity=0.5,->] (2.7,1.1) -- (2.7 ,0.8);
 
 \begin{scope}[yshift=-2.05cm,xshift=-0.6cm,scale=0.82]
	\foreach \x in {0,1} 
	\foreach \y in {0,1} 
	\fill [black,opacity=1] (\x,\y) circle (0.05);
	
	\def\dtog{0.5}
	\draw [densely dashed,opacity=0.6] (0,0) -- (0,-\dtog);
	\draw [densely dashed,opacity=0.6] (1,0) -- (1,-\dtog);
	\draw [<->,opacity=0.6,>=latex]  (0,-\dtog) --(1,-\dtog);
	\node at (0.5,-\dtog) [anchor=north] {$\frac{1}{B}$};
	
	\draw [densely dashed,opacity=0.6] (0,0) -- (-\dtog,0);
	\draw [densely dashed,opacity=0.6] (0,1) -- (-\dtog,1);
	\draw [<->,opacity=0.6,>=latex]  (-\dtog,0) --(-\dtog,1);
	\node at (-\dtog,0.5) [anchor=east] {$\frac{1}{T}$};
	
	\begin{scope}[xshift = 2cm]
	\foreach \x in {0,0.2,...,1} 
	\foreach \y in {0,0.2,...,1} 
	\fill [black,opacity=1] (\x,\y) circle (0.03);
	\end{scope}
	
	\begin{scope}[xshift = 4.5cm]
	\foreach \x in {0,0.1,...,1.001} 
	\foreach \y in {0,0.1,...,1.001} 
	\fill [black,opacity=1] (\x,\y) circle (0.03);
	\end{scope}
	
	\begin{scope}[xshift = 7cm]
	\foreach \x in {0,0.06666,...,1.001} 
	\foreach \y in {0,0.06666,...,1.001} 
	\fill [black,opacity=1] (\x,\y) circle (0.01);
	\end{scope}

 
 \end{scope}
\end{tikzpicture}
 \end{center}
 \vspace{-0.5cm}
 \caption{\label{fig:realrecov} 
Super-resolution radar uniformly provides better resolution error than standard or CS Radar. 
The plot show the resolution error for the recovery of $\S = 10$ time-frequency shifts from the observation $\vy$ with and without additive Gaussian noise $\vn$ of a certain signal-to-noise ratio $\text{SNR} \defeq \norm[2]{\vy}/\norm[2]{\vn}$. 
 The resolution error is defined as the average over  $\L \sqrt{ (\hat \tau_j - \tau_j)^2 + (\hat \nu_j - \nu_j)^2}$, where $(\tau_j,\nu_j)$ are the original time-frequency shifts, and the $(\hat \tau_j, \hat \nu_j)$ are the time-frequency shifts on the grid, obtained  by solving \text{L1-ERR}, for different super-resolution factors. The different super-resolution factors are illustrated below the plot.
 }
\end{figure}


\section{\label{sec:proofoutline}Proof outline}

In this section, we discuss the proofs of Theorems~\ref{thm:mainres} and~\ref{cor:discretesuperres} from~\cite{heckel_super-resolution_2015} which are closely linked. Specifically, both Theorems~\ref{thm:mainres} and~\ref{cor:discretesuperres} follow from the existence of certain dual certificates, and the certificate for proving Theorem~\ref{cor:discretesuperres} is obtained directly from the certificate constructed to prove Theorem~\ref{thm:mainres}. 

\subsection{Proof of Theorem~\ref{thm:mainres}}

Theorem~\ref{thm:mainres} is proven by constructing an appropriate dual certificate; the existence of this  certificate guarantees that the solution to $\AN(\vy)$ is $\vz$, as formalized by 
Proposition~\ref{prop:dualmin} below. 
Proposition~\ref{prop:dualmin} is a consequence of strong duality, and is well known for the discrete setting from the compressed sensing literature \cite{candes_robust_2006}. 
The proof is standard, see for example \cite[Proof of Prop.~2.4]{tang_compressed_2013}. 
For convenience, in this section, we set $\mA = \mG_{\vx}\herm{\mF}$, so that $\vy = \mA \vz$. 

\begin{proposition} 
Let $\vy=\mAA \vz$ with $\vz =   \sum_{j=1}^{\S}  b_j  \atom(\vr_j)$. Suppose there exists a function, called dual certificate, of the form 
$
Q(\vr) =  \innerprod{\vq}{ \mAA \atom(\vr) }
$, parameterized by the complex coefficients $\vq \in \complexset^{\L}$, such that 
\begin{align}
&Q(\vr_j) = \sign(b_j), \text{ for all $j$, and } \nonumber \\
&|Q(\vr)| < 1 \text{ for all } \vr \in [0,1]^2 \setminus \{\vr_1,\hdots,\vr_{\S}\}.
\label{eq:dualpolyinatmincon}
\end{align}
Moreover, suppose that the vectors $\{\mA \vf(\vr_j)\}_{j=1}^{\S}$, are linearly independent. Then $\vz$ is the unique minimizer of $\AN(\vy)$. 
\label{prop:dualmin}
\end{proposition}

A condensed argument showing that Proposition~\ref{prop:dualmin} is true is the following. 
Suppose for contradiction that there exists another optimal solution $\vz' = \sum_{j} b_j' \vf(\vr_j')$ with $\norm[\setA]{\vz'} = \sum_{j} |b_j'|$ and $\{\vr_j'\} \neq \{\vr_j\}$. 
First suppose that  $\{\vr_j'\} \subseteq \{\vr_j\}$. Then, linear independence of the vectors $\{\mA \vf(\vr_j)\}$ contradicts that $\vz' \neq \vz$.
Next, suppose that $\{\vr_j'\} \not\subseteq \{\vr_j\}$.
We then have that
\begin{align*}
\norm[\setA]{\vz'}
-
\norm[\setA]{\vz}
=
\sum_j |b_j'| - \sum_j |b_j|
&\stackrel{i}{>}
\mathrm{Re} \sum_j Q^\ast(\vr_j')b_j'  -  \mathrm{Re} \sum_j Q^\ast(\vr_j)b_j \\ 
&\stackrel{ii}{=}
\mathrm{Re} \sum_j  \herm{\vq}  \mA \vf(\vr_j') b_j' -
\mathrm{Re} \sum_j  \herm{\vq}  \mA \vf(\vr_j) b_j \\
&\stackrel{iii}{=}  0.
\end{align*}
Here, inequality~$i$) follows from the dual polynomial interpolating the sign pattern, and from $|Q(\vr'_j)|<1$ for at least one $\vr_j'$, which in turn follows from $\{\vr_j'\} \not\subseteq \{\vr_j\}$ and $|Q(\vr)|<1$ for all $\vr \notin \{\vr_j\}$, by assumption. 
Inequality~$ii$) follows from the definition of the dual polynomial, and inequality~$iii$) follows from $\mA \vz' = \mA \vz$, by assumption. 
This contradicts that $\vz'$ is an optimal solution.

We now turn to the construction of a dual certificate obeying the conditions of Proposition~\ref{prop:dualmin}, which concludes the proof of Theorem~\ref{thm:mainres}. 
The construction of the dual certificate $Q$ is inspired by the construction of related certificates in \cite{candes_towards_2014,tang_compressed_2013}.
First, recall that the entries of $\atom(\vr), \vr = [\tau,\nu]$, are given by $[\atom(\vr)]_{(k,p)} = e^{i2\pi(k\tau + p\nu)}$. 
From 
\[
Q(\vr) 
= \innerprod{\vq}{ \mAA \atom(\vr)}
= \innerprod{\herm{\mAA}\vq}{\atom(\vr)},
\] 
it is seen that $Q$ is a two-dimensional trigonometric polynomial in the variables $\tau$ and $\nu$ with coefficient vector $\herm{\mA} \vq$. 
To build the certificate, we therefore need to 
construct a two-dimensional trigonometric polynomial that satisfies the interpolation and boundedness condition~\eqref{eq:dualpolyinatmincon}, and whose coefficients are constraint to be of the form $\herm{\mA} \vq$. 
Since the matrix $\mA$ is a function of the random probing signal $\vx$, $Q$ is a \emph{random} trigonometric polynomial. 
We construct $Q$ explicitly. 

It is instructive to first consider the construction of a \emph{deterministic} two-dimensional trigonometric polynomial 
\[
\bar Q(\vr) = \big<\bar \vq,\atom(\vr)\big>
\] 
with unconstraint, deterministic coefficients $\bar \vq \in \complexset^\L$, 
that satisfies the interpolation and boundedness conditions~\eqref{eq:dualpolyinatmincon}, but whose coefficient vector $\bar \vq$ is \emph{not} constraint to be of the form $\herm{\mA} \vq$. 
Such a construction has been established---provided the the parameters $\{\vr_k\}$ obey the minimum separation condition in Definition~\ref{eq:minsepcond}---by Cand\`es and Fernandez-Granda \cite[Prop.~2.1, Prop.~C.1]{candes_towards_2014} for the one- and two-dimensional case. 
To construct the polynomial $\bar Q$, \cite{candes_towards_2014} interpolate the signs $\{\sign(b_j)\}$ with a fast-decaying kernel 
\[
\bar G(\vr) \defeq F(\tau) F(\nu),
\]
and slightly adopt this interpolation near the parameters $\{\vr_j\}$ with the partial derivatives $\bar G^{(n_1,n_2)}(\vr) \defeq 
\frac{\derd^{n_1} }{ \derd \tau^{n_1}} \frac{\derd^{n_2} }{ \derd \nu^{n_2}}  
\bar G(\vr)$ to ensure that local maxima are achieved at the $\vr_j$: 
\begin{align}
\bar Q(\vr) 
= \sum_{j=1}^\S & \bar \alpha_j \bar G(\vr-\vr_j) + \bar \alpha_{1j} \bar G^{(1,0)}(\vr - \vr_j) 
+   \bar\alpha_{2j} \bar G^{(0,1)}(\vr - \vr_j).
\label{eq:detintpolC}
\end{align}
Here, $\FK$ is the squared Fej\'er kernel which is a particular trigonometric polynomial with coefficients $g_j$, i.e.,  
\[
 \FK(t) =  \sum_{j=-\N}^{\N}  g_j e^{i2\pi t j}.
\]
Shifted versions of the polynomial $F$ (i.e., $F(t - t_0)$ for some $t_0 \in \reals$) and the derivatives of the polynomial $\FK$ are also one-dimensional trigonometric polynomials of degree $\N$. 
Therefore $\bar G$, its partial derivatives, and shifted versions thereof are two-dimensional trigonometric polynomials of the form $\innerprod{\bar \vq}{ \atom(\vr)}$. 
The construction of $\bar Q$ is concluded by showing that the coefficients $\bar \alpha_j,\bar \alpha_{1j},\bar \alpha_{2j},\bar \alpha_{3j}$, can be chosen such that $\bar Q$ reaches global maxima at the parameters $\{\vr_j\}$. 

The construction of $Q$ in the paper~\cite{heckel_super-resolution_2015} for proving Theorems~\ref{prop:dualmin} follows a similar program.  
Specifically, the polynomial $Q$ is constructed such that it interpolates the signs $\sign(b_j)$ at $\vr_j$ with the functions 
$
G_{\vn}(\vr,\vr_j) = 
\big<
\mA \vg_{\vn}(\vr_j) ,
\mA \atom(\vr) 
\big>. 
$
Here, $\vg_{\vn}(\vr), \vn = (n_1,n_2)$ is the vector with $(v,k)$-th coefficient given by
\[
g_k g_p
(i2\pi k)^{n_1}(i2\pi p)^{n_2}
e^{-i2\pi (\tau k + \nu p)}, 
\]
where the $\{g_j\}$ are the coefficients of the squared Fej\'er kernel $F$ defined above. 
With this definition, we have 
\[
\EX{G_{\vn}(\vr, \vr_j)} 
=  \bar G^{\vn}(\vr-\vr_j).
\]
This follows from 
\[
\EX{\herm{\mA} \mA } 
=
\herm{\mF} \EX{ \herm{\mG}_\vx \mG_{\vx} } \mF = \mI,
\]
where we used $\EX{ \herm{\mG}_\vx \mG_{\vx} } = \mI$, shown at the end of this section. 
Moreover, $G_{\vn}(\vr, \vr_k)$ concentrates around $\bar G^{\vn}(\vr-\vr_k)$. 

Now, $Q$ is constructed by interpolating the signs $\sign(b_j)$ at $\vr_j$ with $G_{(0,0)}(\vr,\vr_j), j=1,\hdots,\S$, 
and slightly adopting this interpolation near the points $\{\vr_j\}$ with linear combinations of the functions $G_{(1,0)}(\vr,\vr_j)$ and $G_{(0,1)}(\vr,\vr_j)$, in order to ensure that local maxima of $Q$ are achieved exactly at the $\vr_j$. 
Specifically, we set 
\begin{align}
Q(\vr) = \sum_{j=1}^\S 
& \alpha_j G_{(0,0)}(\vr,\vr_j)
+ \alpha_{1j} G_{(1,0)}(\vr,\vr_j) 
+ \alpha_{2j} G_{(0,1)}(\vr,\vr_j).
\label{eq:dualpolyorig}
\end{align}  
Note that $Q(\vr)$ is a linear combination of the functions $G_{\vm}(\vr,\vr_j)$, and by definition of  $G_{\vm}(\vr,\vr_j)$ it obeys $\innerprod{\herm{\mA}\vq}{\atom(\vr)}$, for some $\vq$, as desired. 
The proof is concluded by showing that, with high probability, there exists a choice of coefficients $\alpha_j, \alpha_{1j}$, and $\alpha_{2j}$ such that $Q$ reaches global maxima at the $\vr_j$ and $Q(\vr_j) = u_j$, for all $j$. 
For this argument to work, the particular choice of $G_{\vm}(\vr, \vr_j)$ is crucial; the main ingredients for the argument to work are that $G_{\vm}(\vr, \vr_j)$ concentrates around $\bar G(\vr - \vr_j)$, and certain properties of the deterministic functions $\bar G$ and $\bar Q$.  

\paragraph{Proof of $\EX{\herm{\mG_\vx} \mG_\vx} = \mI$:}
By definition of the Gabor matrix in \eqref{eq:defgabormtx}, the entry in the $(k,\ell)$-th row and $(k',\ell')$-th column of $\herm{\mG}_\vx \mG_\vx$ is given by 
\[
[\herm{\mG}_\vx \mG_\vx]_{(k,\ell), (k',\ell')}  =  \sum_{p=-\N}^\N \conj{x}_{p-\ell} x_{p-\ell'} e^{-i2\pi \frac{kp}{\L}}  e^{i2\pi \frac{k'p}{\L}}.
\]
Noting that $\EX{x_\ell} = 0$, we conclude that $\EX{[\herm{\mG}_\vx \mG_\vx]_{(k,\ell), (k',\ell')}} = 0$ for $\ell \neq \ell'$. For $\ell = \ell'$, using the fact that $\EX{\conj{x}_{p-\ell} x_{p-\ell}} = 1/\L$, we arrive at 
\[
\EX{ [\herm{\mG}_\vx \mG_\vx]_{(k,\ell), (k',\ell')} } = \frac{1}{\L} \sum_{p=-\N}^\N    e^{i2\pi \frac{(k' - k )p}{\L}}.
\]
The latter is equal to $1$ for $k = k'$ and $0$ otherwise. This concludes the proof of $\EX{\herm{\mG}_\vx \mG_\vx} = \mI$. 


\subsection{\label{sec:proofoutline}Proof of Theorem~\ref{cor:discretesuperres}}

The following proposition, which is standard in the compressed sensing literature (see e.g.,~\cite{candes_robust_2006}) states that the existence of a dual polynomial guarantees that $\mathrm{L1}(\vy)$ recovers $\vb$ from the measurement $\vy = \mG \vb$. 
The proposition is the discrete analogue of Proposition~\ref{prop:dualmin} above. 

\begin{proposition}
Let $\vy=\mR\vb$ and let $\dT$ denote the support (i.e., the set of non-zero elements) of $\vb$. Assume that the columms of $\mR$ corresponding to $\dT$ are linearly independent. If there exists a vector $\vv$ in the row space of $\mR$ with 
\begin{align}
\vv_{\dT} = \sign(\vb_{\dT}) \quad \text{and} \quad \norm[\infty]{\vv_{\comp{\dT}}} < 1,
\label{eq:dualcerl1}
\end{align}
then $\vb$ is the unique minimizer of $\mathrm{L1}(\vy)$. 
Here, $\vv_\dT$ its the vector consisting of the entries of $\vv$ indexed by $\dT$, and likewise $\vv_{\comp{\dT}}$ consists of the entries not indexed by $\dT$, i.e., the entries indexed by the complement of $\dT$.
\end{proposition}

Theorem~\ref{cor:discretesuperres} now follows directly from the existence of the polynomial $Q$ constructed in the previous section. To see this, define $\vv$ as
 $[\vv]_{(m,n)} = Q([m/\K,n/\K])$ and note that $\vv$ satisfies \eqref{eq:dualcerl1} since $Q([m/\K,n/\K]) = \sign(\vb_{(m,n)})$ for $(m,n) \in \dT$ and $\abs{Q([m/\K,n/\K])} < 1$ for $(m,n) \notin \dT$.

\section{\label{sec:MIMO}MIMO\index{MIMO}\index{multiple input multiple output} radar}

In this section, we discuss super-resolution imaging in the context of MIMO radar. A MIMO radar uses multiple transmit antennas to send---typically orthogonal---probing signals simultaneously, and records the reflections from the objects with multiple receive antennas. 
As shown in this section, a MIMO radar can thereby, in principle, resolve the relative angles in addition to the relative distances and velocities of objects with a single measurement. 

\begin{figure}[h!]
\begin{center}
\usetikzlibrary{intersections,decorations.pathreplacing}
\begin{tikzpicture}[scale=0.7]
\node (r1) at (0,0) [shape=circle,draw,inner sep =0.1cm] {};
\node (r2) at (0,-1) [shape=circle,draw,inner sep =0.1cm] {};
\node (r3) at (0,-2) [shape=circle,draw,inner sep =0.1cm] {};

\node at (0,1) [] {$\times$};

\foreach \y in {1} {
\draw [domain=-35:35] plot ({1*cos(\x)}, {\y+sin(\x)});
\draw [domain=-35:35] plot ({2*cos(\x)}, {\y+2*sin(\x)});
\draw [domain=-35:35] plot ({3*cos(\x)}, {\y+3*sin(\x)});
};

\node (object) at (10,-2) [fill,shape = rectangle,draw,inner sep=0.1cm]{};

\path[color=gray,draw,dashed,<-] (r1) -- (object);
\path[color=gray,draw,dashed,<-] (r2) -- (5,-2);
\path[color=gray,draw,dashed,<-] (r3) -- (5,-3);
\node at (10,-1) {object};

\path[color=gray,draw,dashed] (r1) -- (4,0);

\node at (3.7,-0.4) {$\theta$};
\draw[gray,draw,dashed] (4,0) arc(0:-12:4);

\draw [decorate,decoration={brace,amplitude=3pt},xshift=-4pt,yshift=0pt]
(-0.2,-2) -- (-0.2,-1) node [black,midway,xshift=-0.4cm] 
{\footnotesize $d_R$};
\end{tikzpicture}
\end{center}
\caption[Principle of MISO radar]{ \label{fig:geomsisorad}
Principle of MISO radar: $\times$ and \tikz \node (r1) at (0,-0.05cm) [shape=circle,draw,inner sep =0.1cm] {}; correspond to transmit  and receive antennas. 
The transmit antenna sends a probing signal, and the reflections of the object are received by three receive antennas. Estimating the relative delays of the probing signal allows to estimate the angle of the object relative to the antenna array, in addition to distance and velocity.
}
\end{figure}
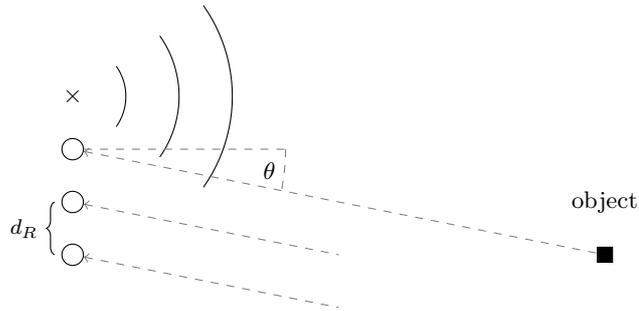

To illustrate the principle of a MIMO radar, first consider a radar system with a single transmit and multiple receive antennas, and consider a object that lies in the far field of the radar, so that the reflections of the objects that arrive at the receiver are essentially parallel, as illustrated in Figure~\ref{fig:geomsisorad}. 
The reflection from the object must travel an additional distance of $d_R \sin(\theta)$ between the signals received at two adjacent receive antennas. 
Thus, from an estimate of the angle of arrival we can determine the relative position of a object. 
The angular resolution that can be achieved as well as the number of objects that can be distinguished increases linearly in the number of receive antennas.  

As a consequence, for doubling the angular resolution or doubling the number of objects to be distinguishable, a MISO radar need to double its number of transmit antennas.
As we discuss next, using multiple transmit antennas in addition to multiple receive antennas can give a much larger angular resolution from far fewer antennas. 
Specifically, by arranging $N_T$ transmit and $N_R$ receive antennas in a particular way (see Figure~\ref{fig:geomrad}) a MIMO radar can obtain the same resolution obtained by a MISO (or SIMO) radar with $N_T N_R$ uniformly spaced receive (or transmit) antennas. This is often called a MIMO \emph{virtual array}. 
In this section, we discuss a MIMO radar model, and show that the fundamental limit for resolving the angle-delay-Doppler triplets is $(1/(N_TN_R), 1/B, 1/T)$.
We furthermore show that this limit can be overcome in the sense that triplets can be resolved on a much finer grid, provided they are sufficiently separated. 


\subsection{MIMO Signal model and problem statement}

\begin{figure}[h!]
\begin{center}
\usetikzlibrary{intersections,decorations.pathreplacing}
\begin{tikzpicture}[scale=0.65]
\node (r1) at (0,0) [shape=circle,draw,inner sep =0.1cm] {};
\node (r2) at (0,3) [shape=circle,draw,inner sep =0.1cm] {};
\node at (0,6) [shape=circle,draw,inner sep =0.1cm] {};

\node at (0,0) [] {$\times$};
\node at (0,-1) [] {$\times$};
\node at (0,-2) [] {$\times$};

\foreach \y in {0,-1,-2} {
\draw [domain=-35:35] plot ({1*cos(\x)}, {\y+sin(\x)});
\draw [domain=-35:35] plot ({2*cos(\x)}, {\y+2*sin(\x)});
\draw [domain=-35:35] plot ({3*cos(\x)}, {\y+3*sin(\x)});
};

\node (object) at (10,-2) [fill,shape = rectangle,draw,inner sep=0.1cm]{};

\node (object3) at (8,0.5) [gray, fill,shape = rectangle,draw,inner sep=0.1cm]{};
\node (object4) at (9,2) [gray,fill,shape = rectangle,draw,inner sep=0.1cm]{};

\path[color=gray,draw,dashed] (r1) -- (object);
\node at (10,-1) {object 1};

\path[color=gray,draw,dashed] (r1) -- (4,0);

\node at (3.7,-0.4) {$\theta$};
\draw[gray,draw,dashed] (4,0) arc(0:-12:4);

\draw [decorate,decoration={brace,amplitude=3pt},xshift=-4pt,yshift=0pt]
(-0.2,-2) -- (-0.2,-1) node [black,midway,xshift=-0.4cm] 
{\footnotesize $d_T$};

\draw [decorate,decoration={brace,amplitude=3pt},xshift=-4pt,yshift=0pt]
(-0.2,3) -- (-0.2,6) node [black,midway,xshift=-0.4cm] 
{\footnotesize $d_R$};

\draw [decorate,decoration={brace,amplitude=3pt},xshift=-4pt,yshift=0pt]
(10.1,-2.3) -- (0.08,-0.3) node [black,midway,yshift=-0.3cm,xshift=-0.1cm] 
{\footnotesize $d$};

\node (rft) at (4.5,5.3) {reflection from object 1};
\draw[->] (2,5) to [out=-90,in=0] (1,4);

\draw[] (r1) -- (1.2,6);

\node at (-2,0) {$r=0,j=0$};

\end{tikzpicture}
\end{center}
\caption[Principle of MIMO radar]{ \label{fig:geomrad}
Principle of MIMO radar: $\times$ and \tikz \node (r1) at (0,-0.05cm) [shape=circle,draw,inner sep =0.1cm] {}; correspond to transmit  and receive antennas. 
Throughout, we assume the spacing of the $N_T$ transmit and $N_R$ receive antennas to be $d_T= \frac{c}{2f_c}$ and $d_R= \frac{cN_T}{2f_c}$, where $f_c$ is the carrier frequency. 
}
\end{figure}
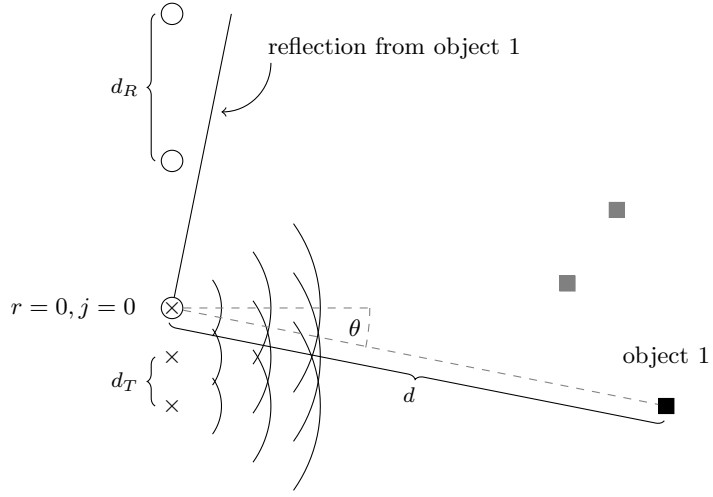

We consider a MIMO radar with $N_T$ transmit and $N_R$ receive antennas that are co-located and lie in a plane along with $S$ objects, see Figure~\ref{fig:geomrad} for an illustration. 
The technical results presented in this section generalize to the more general setup where objects lie in three-dimensional space and the transmit and receive antennas lie in a two-dimensional plane. 
We consider the simpler two-dimensional setup since the generalization to three dimensions are straightforward. 
As in the previous section, we assume that the objects are located in the far field of the array. 
As a consequence, propagating waves appear planar and the angles between the object and each antenna are (approximately) the same. 
We let the transmit and receive antennas be uniformly spaced with spacings $d_T = \frac{1}{2 f_c}$ and $d_R = \frac{N_T}{ 2 f_c}$, respectively, where $f_c$ is the  carrier frequency of the probing signals. 
This spacing yields a uniformly spaced \emph{virtual array} with $N_T N_R$ antennas, and thus maximizes the number of virtual antennas achievable with $N_T$ transmit and $N_R$ receive  antennas \cite{friedlander_relationship_2009,strohmer_analysis_2014}. 
As explained in Section~\ref{sec:iorelintromimo} below, the (baseband) signal $y_r(t)$ at continuous time $t$ received by antenna $r=0,\hdots, N_R-1$, consists of the superposition of the reflections from the objects of the transmitted probing signals $x_j(t), j= 0,\hdots,N_T-1$, and is given by
\begin{align}
y_r(t)
=
\sum_{k=1}^{\S}
b_k 
e^{i2\pi r N_T \beta_k}
\sum_{j=0}^{N_T-1} 
 e^{i2\pi j \beta_k }  x_j(t - \tauc_k) e^{i2\pi \nuc_k t}.
 \label{eq:iorelintromimo}
\end{align}
Here, $b_k \in \complexset$ is the attenuation factor, $\beta_k \in [0,1]$ the angle or azimuth parameter, and $\tauc_k$ and $\nuc_k$ are the delay and Doppler shift, all associated with the $k$-th object. 
The parameters $\beta_k, \tauc_k,\nuc_k$ determine the angle ($\beta = -\sin(\theta)/2$ see Figure~\ref{fig:geomrad}), distance, and velocity of the $k$-th object relative to the radar. 
Locating the object therefore amounts to estimating the continuous parameters $b_k,\beta_k,\tauc_k,\nuc_k$ 
from the responses $y_r, r=0,\hdots,N_R-1$, to known and suitably selected probing signals $x_j$. 

As discussed in Section~\ref{sec:siso}, due to practical constraints, the probing signals must be band-limited and approximately time-limited, and the responses to the probing signals can only be observed over a finite time interval. 
We make the same assumption on the probing signals as well as on the received signals as we made in Section~\ref{sec:siso}; in particular we assume that the received signals $y_r$ are observed over an interval of length $T$ and that the probing signals $x_j$ have bandwidth $B$ and are approximately supported on a time interval proportional to $T$. 
As explained in Section~\ref{sec:siso}, it follows from the band-limitation of the probing signals $x_t$ and the limited observation time of the received signals $y_r$ that the received signals are characterized by the samples 
\begin{align}
[\vy_r]_p
&= 
\sum_{k=1}^{\S}
b_k 
e^{i2\pi r N_T \beta_k } 
\sum_{j=0}^{N_T-1}
e^{i2\pi j \beta_k } 
[\mc F_{\nu_k}
\mc T_{\tau_k}
\vx_j ]_p.
\label{eq:periorelmimo}
\end{align}
Here, $\vy_r$ contains the samples of the received signal $y_r$ taken at rate $1/B$ (i.e., $[\vy_r]_p = y_r(p/B)$ in the interval $ p/B \in [-T/2, T/2]$), and $\vx_j$ is the vector containing the samples of the probing signal ($[\vx_j]_p \defeq x_j(p/B)$). 

We have reduced the problem of identifying the locations of the objects 
under the constraints that the probing signals $x_j$ are band-limited and the responses $y_r$ are observed over a finite time interval only, 
to the estimation of the parameters $b_k \in \complexset$, $(\beta_k,\tau_k, \nu_k )\in [0,1]^3, k=1,\hdots,\S$, from the samples $[\vy_r]_p, r = 0,\hdots,N_R-1, p=-\N,\hdots,\N$, in the input-output relation~\eqref{eq:periorelmimo}. 
We call this the super-resolution MIMO radar problem. 


\subsection{ \label{sec:iorelintromimo} Derivation of the MIMO input-output relation}

In this section, we derive the MIMO input-output relation~\eqref{eq:iorelintromimo}. 
Towards this goal, we first consider a single object. 
The $j$-th antenna transmits the signal $x_j(t) e^{i2\pi f_c t}$, where $f_c$ is the carrier frequency. 
This signal propagates to the object, which we assume to be a point scatterer, gets reflected, and propagates back to the $r$-th receiver. 
From Figure~\ref{fig:geomrad}, we see that the corresponding delay is, as a function of the angle between antennas and the object, $\theta$,  distance to the object, $d$, and speed of light, $c$, given by
\[
\tilde \tau 
\defeq
\frac{ 2d }{c}
+
\frac{\sin(\theta)(d_T j + d_R r )}{c}
= 
 \tauc - \beta  \frac{2(d_T j + d_R r ) }{c}. 
\]
For the second equality, we defined the angle parameter $\beta \defeq -\sin(\theta)/2$ and the delay $\tauc \defeq \frac{2d}{c}$. 
Taking the Doppler shift into account, the reflection of the $j$-th probing signal 
received by the $r$-th receive antenna is given by 
\begin{align}
\tilde b x_j(t   - \tilde \tau ) 
e^{i2\pi (f_c + \nuc ) \left( t -  \tilde \tau \right)}.
\label{eq:exresp}
\end{align}
Here, $\tilde b\in \complexset$ is the attenuation factor associated with the object, and $\nuc \defeq \frac{2 v }{c} f_c$ is the Doppler shift, which is a function of the relative velocity, $v$, of the object. 
By choosing the antenna spacing as $d_T = \frac{c}{2f_c}$ and $d_R =\frac{cN_T}{2f_c}$, 
the reflection of the $j$-th probing signal received by the $r$-th receive antenna in \eqref{eq:exresp} becomes 
\[
\tilde b x_j( t - \tilde \tau ) 
e^{i2\pi (f_c + \nuc) \left( t -  \tauc \right)}
e^{i2\pi (f_c + \nuc) 
\beta \frac{j + r N_T}{f_c}       
}
\approx 
\tilde b x_j(t   - \tauc) 
e^{i2\pi (f_c + \nuc) \left( t -  \tauc \right)}
e^{i2\pi  
\beta  (j  + r N_T )   
}.
\]
Here, the approximation follows by the Doppler shift $\nuc$ being much smaller than the carrier frequency $f_c$, therefore $\frac{f_c + \nuc}{f_c} \approx 1$, and 
$\tilde \tau \approx \tauc$. 
If follows that the 
reflection of the $j$-th probing signal received by the $r$-th receive antenna, after demodulation, is 
\[
b x_j(t   - \tauc) 
e^{i2\pi \nuc t }
e^{i2\pi   \beta  (j  + r N_T ) }
\]
where we defined $b = \tilde b e^{-i2\pi \nuc \tauc}$. 
Next, consider $\S$ objects with parameters \allowbreak$(b_k,\beta_k,\tauc_k,\nuc_k)$. 
Since, for $\S$ objects, the (demodulated) signal $y_r$ received by antenna $r$ consists of the superposition of the reflections of the probing signals $x_j,j=0,\ldots,N_T-1$, transmitted by the transmit antennas, we obtain the input-output relation in equation~\eqref{eq:iorelintromimo} simply by summing over the reflections given by $b_k x_j(t   - \tauc_k) 
e^{i2\pi \nuc_k t }
e^{i2\pi   \beta_k  (j  + r N_T ) }$.


\subsection{MIMO atomic norm minimization}

Recall that our goal is to recover the unknown parameters $(b_k,\beta_k,\tau_k,\nu_k)$ from the measurements $\{\vy_r\}$. 
Towards this goal, we proceed analogously as in Section~\ref{sec:bandtimelim}. 
We start by defining for convenience the vector $\vr \defeq [\beta,\tau,\nu]$, and write the input-output relation \eqref{eq:periorelmimo} in matrix-vector form:
\begin{align}
\vy = \mAA \vz, 
\quad \vz = \sum_{k=1}^{\S} b_k \atom(\vr_k).
\label{eq:syseqinwmimo}
\end{align}
Here  $\vy$ is obtained by stacking the vectors $\transp{\vy}_0,\hdots, \transp{\vy}_{N_R-1}$, 
and the vector $\atom(\vr) \in \complexset^{\L^2 N_TN_R}$ has entries 
\[
[\atom(\vr)]_{(v,k,p)} = e^{i2\pi(v \beta + k\tau + p\nu)}, v=0,\hdots,N_T N_R-1, \quad k,p=-\N,\hdots,\N.
\]
Similarly as before, we use for convenience a three dimensional index to refer to entries of a vector. 
Finally, the matrix $\mAA \in \complexset^{N_R \L \times N_R N_T \L^2}$ is defined as follows. The expression 
\[
w_{r,p} \defeq e^{i2\pi r N_T \beta } 
\sum_{j=0}^{N_T-1}
e^{i2\pi j \beta } 
[\mc F_{\nu}
\mc T_{\tau}
\vx_j ]_p
\]
in equation~\eqref{eq:periorel} can be written as
\[
w_{r,p}
=
\sum_{j=0}^{N_T-1}
\sum_{k = -\N}^{\N} 
a_{p,k,j}
   e^{i2\pi (k \tau + p \nu + (j+N_T r) \beta)},
\]
with 
\[
a_{p,k,j} = \frac{1}{\L} \sum_{\ell = -\N}^{\N}
  [\vx_j]_\ell
    e^{i2\pi (\ell-p) \frac{k}{L} }.
\]
Let $\atom_{p,j} \in \complexset^\L$ be the vector with $k$th entry $[\atom_{p,j}]_k = a_{p,k,j}$, $k=-\N,\hdots,\N$, and let $\mA_j \in \complexset^{\L \times \L^2}$ be the block-diagonal matrix with $\transp{\atom}_{p,j}$ on its $p$th diagonal, $p=-\N,\hdots,\N$. With this notation, $\mAA$ is defined as the block-diagonal matrix with the matrix $[\mA_0,\hdots,\mA_{N_T-1}] \in \complexset^{\L \times N_T \L^2}$ on its diagonal, for all $N_R$ blocks on the diagonal. With this notation, \eqref{eq:periorel} becomes \eqref{eq:syseqinwmimo}.

Similarly as for the SISO radar problem, recovery of the unknowns $b_k,\vr_k = [\beta_k,\tau_k,\nu_k]$ from the measurement $\vz = \sum_{k=1}^{\S} b_k \atom(\vr_k)$ is a 3D line spectral estimation problem that can be solved with standard spectral estimation techniques. 
In order to recover the vector $\vz$ from the measurement $\vy$, we use that $\vz$ is a sparse linear combination of  atoms in the set $\setA \defeq \{ \atom(\vr), \vr \in [0,1]^3\}$, and estimate $\vz$ by solving the basis pursuit type atomic norm minimization problem problem  
\[
\AN(\vy) \colon \;\; \underset{\minlet{\vz}  }{\text{minimize}} \,  \norm[\setA]{\minlet{\vz} } \; \text{ subject to } \; \vy = \mAA \minlet{\vz},
\]
where 
\[
\norm[\setA]{\vz} 
\defeq \inf_{b_k \in \complexset, \vr_k \in [0,1]^3} \left\{ \sum_k |b_k| \colon \vz = \sum_k b_k \atom(\vr_k) \right\}.
\]
To summarize, as for the SISO radar problem, we estimate the parameters $b_k,\vr_k$ from $\vy$ by:
\begin{enumerate}
\item[i] solving $\AN(\vy)$ in order to obtain $\vz$, 
\item[ii] estimating the $\vr_k$ from $\vz$ by solving the corresponding 3D-line spectral estimation problem, and 
\item[iii] solving the linear system of equations 
$
\vy = \sum_{k=0}^{\S-1} b_k \mAA \atom(\vr_k)
$ for the $b_k$. 
\end{enumerate}

\subsection{Recovery guarantees for MIMO atomic norm minimization}

As before, we take the probing signals to be random by choosing its samples, i.e., the entries of the $\vx_j$ as i.i.d.~Gaussian (or sub-Gaussian) zero-mean random variables with variance $1/(N_T \L)$. 
Moreover, we again require a minimum separation condition to be satisfied. 

\begin{definition}[MIMO minimum separation condition] We say the triplets 
$(\beta_j,\tau_j,\nu_j) \in [0,1]^2, j = 1,\ldots,\S$, satisfy the minimum separation condition if for all $j,j'\colon j\neq j'$, 
\begin{align}
|\beta_j - \beta_{j'}| \geq \frac{10}{N_T N_R- 1} \quad \text{or}\quad
 |\tau_j - \tau_{j'}| \geq \frac{5}{\N} 
\quad\text{or}\quad
|\nu_j - \nu_{j'}|  \geq \frac{5}{\N}. 
\label{eq:mimominsepcond}
\end{align}
As before, $|\tau_j - \tau_{j'}|$ is the wrap-around distance on the unit circle. 
\end{definition}

Note that the triplets must \emph{not} be separated in angle, time, \emph{and} frequency simultaneously, for the MIMO minimum separation condition to be satisfied, it is sufficient if they are separated in at least one of those dimensions.

\begin{theorem} 
Assume that the samples of the probing signals $\vx_j, j = 0,\hdots,N_T-1$, are i.i.d. zero-mean Gaussian random variables with variance $1/ (N_T\L)$, and let $\L = 2\N+1 \geq 1024$ and $N_T N_R \geq 1024$. 
Consider a signal where the signs of the attenuation factors $\{b_j\}_{j=1}^{\S}$ are i.i.d.~uniform on $\{-1,1\}$ or the complex unit disc, 
and suppose that the triplets $\{(\beta_j, \tau_j,\nu_j)\}_{j=1}^{\S}$
obey the MIMO minimum separation condition. 
Furthermore, choose $\delta > 0$ and assume that the number of non-zero attenuation factors, $\S$, obeys 
\begin{align}
S \le c \frac{\min(L, N_T N_R)}{\log^3\left(L/\delta \right)},
\label{eq:sleqb}
\end{align}
where $c$ is a numerical constant. Then, with probability at least $1-\delta$, $\vz = \sum_{k=1}^{\S} b_k \atom(\vr_k)$ is the unique minimizer of $\AN(\vy)$, $\vy = \mA \vz$. 
\label{thm:mainresmimo}
\end{theorem}


Theorem~\ref{thm:mainres} guarantees that, with high probability, the attenuation factors and location parameters can be recovered perfectly from the observation $\vy$ by solving a convex program (recall that the parameters $b_k, \vr_k$ can be obtained from $\vz$), provided that the locations $\vr_k = [\beta_k,\tau_k,\nu_k]$ are sufficiently separated in \emph{either} angle, time, or frequency, and provided that the total number of objects satisfies condition \eqref{eq:sleqb}. 
Note that, translated to the physical parameters $\tauc_k,\nuc_k$, the MIMO minimum separation condition becomes: For all $k,k'\colon k\neq k'$,
\[
|\beta_k - \beta_{k'}| \geq \frac{10}{N_T N_R- 1}
\quad \text{or} \quad 
 |\tauc_k - \tauc_{k'}| \geq \frac{10.01}{B}
\quad \text{or} \quad 
|\nuc_k - \nuc_{k'}|  \geq \frac{10.01}{T}.
\]

Theorem~\ref{thm:mainres} is essentially optimal in the number of objects that can be located, since $\S$ can be linear---up to a log-factor---in $\min(\L, N_T N_R)$, and $\S \leq \min(\L, N_T N_R)$ is a necessary condition to uniquely recover the attenuation factors $b_k$ even if the locations $\vr_k$ are known. 
To see this, note that for the linear system of equations  \eqref{eq:syseqinwmimo} to have a unique solution, 
the vectors $\mAA \atom(\vr_k)$ must be linearly independent. 
If $\beta_k=0$, for all $k$, or if $\tau_k=0$ and $\nu_k=0$, for all $k$, the vectors $\mAA \atom(\vr_k), \vr_k = (\beta_k, \tau_k, \nu_k), k=0,\hdots,\S-1$ 
can only be linearly independent provided that $\S \leq \L$ and $\S \leq N_T N_R$, respectively. 
This is seen from 
\[
\mAA \atom(\vr)
=
\begin{bmatrix}
e^{i2\pi 0 \beta }  \sum_{j=0}^{N_T-1} e^{i2\pi j \beta} \mc F_{\nu}
\mc T_{\tau} \vx_j  \\
\vdots  \\
e^{i2\pi N_T (N_R-1) \beta }  \sum_{j=0}^{N_T-1} e^{i2\pi j \beta} \mc F_{\nu}
\mc T_{\tau} \vx_j 
\end{bmatrix}.
\]
We finally note that Theorem~\ref{thm:mainres} is proven by constructing a dual certificate in a similar manner as we constructed the certificate for the SISO in Section~\ref{cor:discretesuperresmimo}.

\subsection{\label{sec:discretesuperesmimo} MIMO super-resolution radar on a fine grid}

As discussed before for the SISO radar setup, a practical approach to estimate the parameters $\vr_k$ from the received signals, is to suppose the angle-time-frequency triplets lie on a \emph{fine} grid, and solve the recovery problem on that grid. 
In general this leads to a gridding error, that, however, decreases as the grid becomes finer. 
We next discuss the corresponding (discrete) sparse signal recovery problem. 

Suppose the parameters $(\beta_k,\tau_k,\nu_k)$ lie on a grid with spacing $(1/K_1,1/K_2,1/K_3)$, where $K_1,K_2,K_3$ are integers obeying $K_1 \geq N_TN_R$, $K_2,K_3 \geq \L = 2 \N +1$. 
With this assumption, the super-resolution MIMO radar problem reduces to the recovery of the sparse (discrete) signal $\vb\in \complexset^{K_1 K_2 K_3}$ 
from the measurement
\[
\vy = \mR \vb,
\]
where $\mR \in \complexset^{N_R \L  \times K_1 K_2 K_3}$ is the matrix with $(n_1,n_2,n_3)$-th column given by 
\[
\mAA \atom(\vr_\vn),
\quad 
\vr_\vn = (n_1/\K_1, n_2/\K_2, n_3/\K_3).
\]
Note that the non-zeros of $\vb$ and its indices correspond to the attenuation factors $b_k$ and the locations $\vr_k$ on the grid. 
A standard approach to the recovery of the sparse signal $\vb$ from the underdetermined linear system of equations $\vy = \mR \vb$ is to solve the following convex program:
\begin{align}
	\mathrm{L1}(\vy) \colon \;\;
\underset{\minlet{\vb}}{\text{minimize}} \; \norm[1]{\minlet{\vb}} \text{ subject to } \vy = \mR \minlet{\vb}.
\label{eq:l1minmGmimo}
\end{align}
Below is the main result for recovery on the fine grid.   

\begin{theorem}  
Assume $\L = 2\N+1 \geq 1024$, $N_T N_R \geq 1024$, and suppose we observe $\vy = \mR \vb$, 
where $\vb$ is a sparse vector with non-zeros indexed by the support set $\mc S \subseteq [K_1]\times [K_2] \times [K_3]$, $[K] \defeq \{0,\ldots,K-1\}$. 
Suppose that those indices satisfy the following minimum separation condition: For all triplets $(n_1,n_2,n_3), (n_1',n_2',n_3') \in \mc S$, 
\[
\frac{|n_{1} - n_{1}'|}{K_1} \geq \frac{10}{N_T N_R- 1} 
\quad \text{or}\quad
\frac{|n_{2} - n_{2}'|}{K_2} \geq \frac{5}{\N}
\quad\text{or}\quad
\frac{|n_{3} - n_{3}'|}{K_3}  \geq \frac{5}{\N}.
\]
Moreover, we assume that the signs of the non-zeros of $\vb$ are chosen independently from symmetric distributions on the complex unit circle. 
Choose $\delta>0$ and assume
\[
S \le c \frac{\min(L, N_T N_R) }{ \log^3\left(\L/\delta \right)},
\]
where $c$ is a numerical constant. Then, with probability at least $1-\delta$, $\mathbf{s}$ is the unique minimizer of $\mathrm{L1}(\vy)$ in \eqref{eq:l1minmGmimo}. 
\label{cor:discretesuperresmimo}
\end{theorem}

Note that Theorem~\ref{cor:discretesuperresmimo} does not impose any restriction on $K_1,K_2,K_3$, in particular they can be arbitrarily large.  
The proof of Theorem~\ref{cor:discretesuperresmimo} is closely linked to that of Theorem~\ref{thm:mainresmimo}; 
specifically, similarly as for the SISO case, the existence of a certain dual certificate guarantees that $\vb$ is the unique minimizer of $\mathrm{L1}(\vy)$. 
The dual certificate is obtained directly from the dual certificate for the continuous case, which, as mentioned before, is constructed for the MIMO case in a similar way as the certificate for the SISO case has been constructed in Section~\ref{cor:discretesuperresmimo}. 


\subsection{\label{sec:numresmimo} Numerical results and robustness}

Paralleling the discussion in Section~\ref{sec:numres} for  SISO radar, we next briefly numerically evaluate the resolution obtained by the norm minimization approach to enable super-resolution in a MIMO radar, and demonstrate robustness to noise.
We discuss a synthetic experiment from~\cite{heckel_super-resolution_2016}. In that experiment, a problem instance was generated by setting $N_T = 3, N_R = 3$, $\L = 41$, and $S=5$ object locations $(\beta_k,\tau_k,\nu_k)$ were drawn uniformly at random from $[0,1] \times [0,2/\sqrt{\L}]^2$. Moreover, we choose $K_1 = \SRF N_T N_R, K_2 = \SRF \L$, and $K_3 = \SRF \L$, where $\SRF\geq 1$ can be interpreted as a super-resolution factor as it determines by how much the $(1/K_1,1/K_2,1/K_3)$ grid is finer than the original, coarse grid $(1/(N_TN_R),1/\L,1/\L)$. 
 To account for additive noise, as before, we solve the following modification of $\mathrm{L1}(\vy)$ in \eqref{eq:l1minmGmimo}
\[
\text{L1-ERR}\colon \underset{\minlet{\vb}}{\text{minimize}} \; \norm[1]{\minlet{\vb}} \text{ subject to } 
\norm[2]{\vy - \mR \minlet{\vb} }^2 \leq \delta,
\]
with $\delta$ chosen on the order of the noise variance. 
There are two error sources incurred by this approach: the gridding error obtained by assuming the points lie on a grid with spacing $(1/K_1,1/K_2,\allowbreak1/K_3)$, which decreases in $\SRF$ and becomes negligible, and the additive noise error, which is constant. 
The results of the simulations, depicted in Figure~\ref{fig:realrecov}, show that the object resolution of the super-resolution approach is significantly better than that of the compressed sensing based approach \cite{dorsch_refined_2015,strohmer_adventures_2015} corresponding to recovery on the coarse grid, i.e., $\SRF=1$. 
Moreover, the results show that the approach is robust to noise and that even under noise, the localization accuracy is significantly improved over a standard approach to radar. 

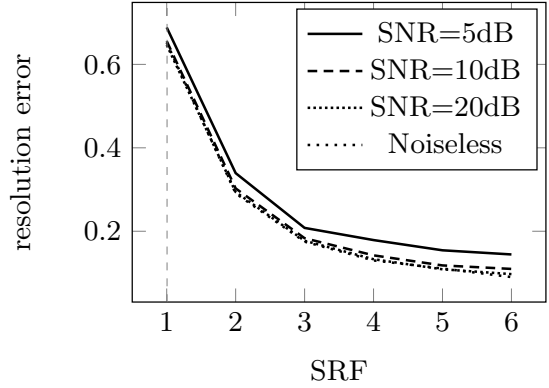
\begin{figure}
\begin{center}
 \begin{tikzpicture}[scale=1.2,thick] 
 \begin{axis}[xlabel={$\text{SRF}$}, ylabel = {resolution error},  xtick={1,2,3,4,5,6},
 width = 0.5\textwidth, height=0.4\textwidth]   
 \addplot +[thick,mark=none,black] table[x index=0,y index=2]{./dat/offgrid.dat};
 \addlegendentry{SNR=5dB}
  \addplot +[thick,mark=none,densely dashed,black] table[x index=0,y index=3]{./dat/offgrid.dat}; 
 \addlegendentry{SNR=10dB}
  \addplot +[thick,mark=none, densely dotted,black] table[x index=0,y index=4]{./dat/offgrid.dat};
 \addlegendentry{SNR=20dB} 
  \addplot +[thick,mark=none, dotted,black] table[x index=0,y index=1]{./dat/offgrid.dat}; 
 \addlegendentry{Noiseless}
  \end{axis}
  \draw[very thin,dashed,gray] (0.38,0) -- (0.38,3.3);
 \end{tikzpicture}
 \end{center}
 \caption{\label{fig:realrecov} 
 Resolution error for the recovery of $\S = 5$ objects from the samples $\mathbf y$ with and without additive Gaussian noise $\mathbf n$ of a certain signal-to-noise ratio $\text{SNR} = \| \mathbf y \|^2_2/ \| \mathbf n \|^2_2$, for varying super-resolution factors (SRFs).  
 The resolution error is defined as the average over $( N_T^2 N_R^2(\hat \beta_k - \beta_k)^2 + \L^2(\hat \tau_k - \tau_k)^2 + \L^2(\hat \nu_k - \nu_k)^2)^{1/2}$, $k=1,\ldots,\S$,  where $(\hat \beta_k,\hat \tau_k, \hat \nu_k)$ are the locations obtained by solving \text{L1-ERR}.   
}
\end{figure}

We next compare our approach to the  Iterative Adaptive Approach (IAA)~\cite{yardibi_source_2010}, proposed for MIMO radar in the paper~\cite{roberts_iterative_2010}. IAA is based on weighted least squares and has been proposed in the array processing literature. IAA  can work well even with one snapshot only and can therefore be directly applied to the MIMO super-resolution problem. 
However, to the best of our knowledge, no analytical performance guarantees are available in the literature that attest IAA similar performance than the $\ell_1$-minimization based approach. 
We compare the IAA algorithm \cite[Table II, ``The IAA-APES Algorithm'']{yardibi_source_2010} to L1-ERR, for a problem with parameters 
$N_T = 3, N_R = 3$, and $\L = 41$, as before, but with $\SRF = 3$ and $(\beta_k,\tau_k,\nu_k) = (k/(N_R N_t), k/\L, k/\L), k=1,\ldots,\S$, so that the 
location parameters lie on the fine grid, and are separated. 
As before, we draw the corresponding attenuation factors $b_k$ i.i.d.~uniformly at random from the complex unit disc. 
Our results, depicted in Figure \ref{fig:cmpmusicconvex}, show that L1-ERR performs better in this experiment than IAA, in particular for small signal-to-noise ratios.

\begin{figure}[!ht]
\centering
\begin{tikzpicture}    
\begin{axis}[
xlabel = SNR in dB,
ylabel=resolution error,
width = 0.5\textwidth,
height=0.4\textwidth,
x dir=reverse,
legend pos=north west,
yticklabel style={/pgf/number format/fixed,
                  /pgf/number format/precision=3}
]

\addplot +[black,mark options={black}] table[x index=0,y index=2]{./dat/resultsSRF_nfp.dat};
\addlegendentry{IAA};

\addplot +[black,mark options={black}] table[x index=0,y index=1]{./dat/resultsSRF_nfp.dat};
\addlegendentry{ L1-ERR };

\end{axis}
\end{tikzpicture}

\caption{
\label{fig:cmpmusicconvex}
Resolution error (smaller is better) of L1-ERR and IAA applied to $\vy + \vn$, where $\vn \in \complexset^{N_R\L}$ is additive Gaussian noise, such that the signal-to-noise ratio is 
 $\text{SNR} \defeq \norm[2]{\vy}^2 / \norm[2]{ \vn }^2$. 
 As before, the resolution error is defined as $( N_T^2 N_R^2(\hat \beta_k - \beta_k)^2 + \L^2(\hat \tau_k - \tau_k)^2 + \L^2(\hat \nu_k - \nu_k)^2)^{1/2}$, 
where $(\hat \beta_k,\hat \tau_k, \hat \nu_k)$ are the locations obtained by solving \text{L1-ERR}.  
}

\end{figure} 


\section{\label{sec:discuss}Discussion and current and future research directions}

In this section we discuss a class of signal recovery problems that are closely related to the SISO and MIMO radar problem, corresponding results and open theoretical research problems, and comment on computational challenges in applying the methods discussed here in practical radar systems.

The SISO and MIMO radar problems discussed in this chapter are versions of a more general problem, namely that of recovering a signal that is sparse in a \emph{continuously} indexed dictionary, with the index corresponding to the locations and velocities of objects. 
In contrast, traditional compressive sensing research has focused on the recovery of signals that are sparse in \emph{discretely} indexed dictionaries via convex programs~\cite{candes_robust_2006} amongst other methods. 
As discussed in this chapter in the context of radar, signals that are sparse in continuously indexed dictionaries can be recovered via a convex program either by solving an atomic norm minimization problem, or by discretizing the continuous parameter space. However, the discretization step induces a gridding error. While in practice---provided the grid is chosen sufficiently fine---the gridding error is negligible, fine discretization leads to dictionaries with extremely correlated, i.e., coherent, columns, and the theory of compressive sensing, and many practical algorithms, rely on the dictionary to be incoherent and therefore does not apply to fine grids.
The primary difficulty with recovering signals in such dictionaries is that the elements of the dictionaries are very close to each other---which is the case both for continuously indexed dictionaries as well as for finely discretized signals. 
Stable recovery of signals that are sparse in such dictionaries requires excluding signals that are supported on elements of the dictionary that are very close to each other. 
Such signals are excluded here by imposing the minimum separation condition. 


More specifically, the SISO and MIMO radar problems belong to a class of signal recovery problems where the goal is to recover unknown coefficients $\{b_j\}$ and location parameters $\{\vr_j\}$ from the measurement
\[
\vy = \sum_{j=1}^\S b_j \mA \vf(\vr_j).
\]
Here, $\vf(\vr)$ is a vector containing complex exponentials, and $\vr$ is a $d$-dimensional location parameter. For example, if $\vr$ is a one-dimensional location parameter then $[\vf(\vr)]_{r} = e^{-i2\pi r\tau}$, and if $\vr$ is two-dimensional location vector, as in the SISO radar problem, then $[\vf(\vr)]_{(r,q)} = e^{-i2\pi (r\tau + q \nu)}$. The matrix $\mA$ is a problem dependent matrix that parameterizes the dictionary; in the SISO case it is equal to 
$\mA = \mG_{\vx}\herm{\mF}$, see equation~\eqref{eq:syseqinw}. 
In other words, radar signals are sparse in a continuously dictionary that is parameterized by a matrix $\mA$. 
We hasten to add that there are a number of interesting signal recovery problems in continuously indexed dictionaries that are not of this particular form, the deconvolution problem in the paper~\cite{bernstein_deconvolution_2017} is such an example, and the computational imaging problem in~\cite{antipa_diffusercam_2018} is another. 


\subsection{Stability to noise}

In this section, we discuss analytical results on the stability to noise of the atomic norm minimization framework. While currently there are no formal results for the SISO and MIMO radar problem, we discuss statements pertaining to two closely related problems. 
The first is the classical line spectral estimation problem where $\mA$ is the identity matrix, and the second is a generalized line spectral estimation problem, where $\mA$ is a Gaussian random matrix. 

As mentioned before, for $\mA=\mI$, the sparse recovery problem reduces to the classical line spectral estimation problem studied for the noisy and noiseless case in~\cite{candes_towards_2014,candes_super-resolution_2014,fernandez-granda_super-resolution_2015,bhaskar_atomic_2012}.
This problem is well understood, and atomic norm minimization succeeds under very general conditions. 
Specifically, the paper~\cite{candes_towards_2014} shows that a convex program provably recovers the coefficients $\{b_j\}$ and location parameters $\{\vr_j\}$ perfectly, provided the minimum separation condition holds (see Section~\ref{sec:mainares}). 
While standard spectral estimation techniques such as Prony's method, MUSIC, and ESPRIT \cite{stoica_spectral_2005} also provably succeed for the noiseless case, even without requiring a separation condition, an advantage of the convex program is that it does not require knowledge of $\S$, and perhaps more importantly is provably robust~\cite{tang_near_2015,candes_super-resolution_2014,bhaskar_atomic_2012}. 

Specifically, Tang et al.~\cite{tang_near_2015} show that the atomic norm regularized least squares estimate enables near-optimal denoising of $\vz$ from a noisy measurement $\vy = \vz + \ve$, where $\vz$ is a signal of the form $\vz = \sum_{j=1}^\S \vf(\nu_j)$, and $\ve$ is zero-mean Gaussian noise with variance $\sigma^2\mI$. 
Specifically, provided the minimum separation condition holds, the atomic norm regularized least squares estimate $\hat \vz$ obeys 
\[
\norm[2]{\hat \vz  - \vz}^2 \leq c \sigma^2 \S \log(\L),
\]
with high probability. 
This result is essentially optimal; to see this, note that even if we knew the location parameters $\{\nu_j\}$ exactly, the best bound we could achieve would be $\sigma^2 \S$~\cite{bunea_sparsity_2007}, only by a logarithmic factor short of the result reviewed above. 
In addition, the paper~\cite{tang_near_2015} shows that the corresponding estimator localizes the frequencies up to a certain (small) error, provided that the number of samples is sufficiently large.

Next, suppose that $\mA$ is a $M \times L$ Gaussian random matrix with i.i.d.~$\mc N(0,1/M)+i\mc N(0,1/M)$ entires, with $M$ typically much smaller than $L$. 
Assume we are given a noisy measurement 
$\vy = \mA \vz + \ve$, with $\vz$ a signal of the form $\vz = \sum_{j=1}^\S \vf(\nu_j)$ and $\ve$ is noise. 
To estimate the signal $\vz$ from such a measurement one can use an atomic norm optimization problem of the form
\begin{align}
\label{noi}
\hat{\vz}:=\underset{\bar{\vz}}{\arg\min}\text{ }\frac{1}{2}\norm[2]{\vy-\mA\bar{\vz}}^2\quad\text{subject to}\quad\norm[\setA]{\bar{\vz}}\le \tau,
\end{align}
with $\tau$ a tuning parameter. Theorem 4 in the paper~\cite{heckel_generalized_2017} shows that, as long as 
\[
M \geq c S \log(L),
\]
with $c$ a fixed numerical constant, the minimizer of \eqref{noi} with $\tau=\|\vz\|_{\mathcal{A}}$ obeys 
\begin{align*}
\norm[2]{\hat{\vz}-\vz}^2 \le cS\log(L)\sigma^2,
\end{align*}
with high probability. 
Again, this results is essentially optimal both with respect to the number of measurements $M$ required relative to the number of unknowns $S$, as well as with respect to the best bound we can achieve for estimation under noise. 
Moreover, this results do not make any assumptions on the coefficients $b_j$. 
Unfortunately, the corresponding proof strategy does not carry over to the case where $\mA$ is a structured random matrix, as in the SISO and MIMO radar problem considered here. However, numerical simulations suggest that for a number of structured random matrices, including the matrices parametrizing the SISO and MIMO radar problems, the performance of the nuclear norm minimization program as well as $\ell_1$-norm regularized least squares, is similar. 

\subsection{Computational challenges}

A challenge in applying the convex optimization based super-resolution methods in the context of radar is their  computational complexity. 
Specifically, in the context of SISO radar, if we estimate the time-frequency shifts based on solving (the dual of) the atomic norm minimization problem, then the optimization variable of the corresponding convex program has dimensions $\L^2 \times \L^2$. Thus, an algorithm that solves or approximates the atomic norm minimization problem has computational complexity at least $L^4$, which is infeasible for real-world problems.
As discussed in Section~\ref{sec:discretesuperes}, what comes to our rescue is that in practice, we can solve the super-resolution radar problem on a fine grid, and recover the signal by solving a $\ell_1$-minimization problem. 
The complexity of numerically solving the corresponding program with a standard iterative algorithm such as FISTA~\cite{beck_fast_2009} depends on the dimension of the matrix (determined by the problem size ($BT$ and number of antennas)), as well as the corresponding super-resolution factor (see Section~\ref{sec:impdet}), and the conditioning of the matrices involved. 
Increasing the super-resolution factor leads to both a larger problem size (the number of columns in the SISO radar problem increases quadratically in the super-resolution factor), which results in a larger iteration complexity, as well as in general to a slower convergence of the iterative algorithm, since the conditioning of the matrices involved become worse. 
Thus, an interesting research direction is to develop computationally efficient algorithms for the recovery of signals in continuously indexed dictionaries in general, and for the SISO and MIMO radar problem in particular. 
See the papers~\cite{boyd_alternating_2017,rao_forward-backward_2015} for some recent work in this direction.





  \backmatter
 \bibliographystyle{IEEEtran} 

  \bibliography{chap_heckel_refs}
  \label{refs}

\cleardoublepage

\end{document}